\documentclass[aps,prx,groupedaddress,twocolumn,showpacs,longbibliography,10pt]{revtex4-2}

\usepackage{amsmath}
\usepackage{amsthm, amssymb,bbold}    
\usepackage{bm} 
\usepackage{braket}              
\usepackage{graphicx}
\usepackage{color}
\usepackage{times, verbatim}      
\usepackage{mathtools,tikz}
\usetikzlibrary{matrix}
\usepackage{paralist}
\usepackage{natbib}
\usepackage{multirow}
\usepackage{hyperref}
\usepackage{wrapfig,lipsum,booktabs}
\hypersetup{
       colorlinks=true,
}

\newcommand{\CO}{{\cal O}}

\newcommand{\CC}{{\cal C}}

\newcommand{\CZ}{{\cal Z}}

\newcommand{\tSp}{{\widetilde{\text{Sp}}}}

\newcommand{\so}{{\mathfrak{so}}}

\newcommand{\SO}{{\text{SO}}}
\newcommand{\SU}{{\text{SU}}}

\makeatletter
\newcommand*{\rom}[1]{\expandafter\@slowromancap\romannumeral #1@}
\makeatother

\usepackage{color}
\definecolor{darkgreen}{rgb}{0,0.5,0}
\definecolor{darkblue}{rgb}{0,0,0.6}
\definecolor{purple}{rgb}{0.4,.2,0.7}

\def\CH{{\cal H}}  
 
\def\CZ{{\cal Z}}

\def\tr{\,{\rm tr}\,}

\newcommand{\CA}{\mathcal{A}}

\DeclareFontShape{OT1}{cmr}{mx}{n}{<->cmr10}{}

\begin{document}

\fontseries{mx}\selectfont

\title{Majorana Scars as Group Singlets} 

\author{Zimo Sun,$^{1}$ Fedor K. Popov,$^{2}$ Igor R. Klebanov,$^{1,3}$ Kiryl Pakrouski$^{4}$}
\affiliation{$^{1}$Department of Physics, Princeton University, Princeton, NJ 08544, USA}
\affiliation{$^{2}$Department of Physics, New York University, New York, NY 10003, USA}
\affiliation{$^{3}$Princeton Center for Theoretical Science, Princeton University, Princeton, NJ 08544, USA}
\affiliation{$^{4}$Institute for Theoretical Physics, ETH Zurich, 8093 Zurich, Switzerland}

\begin{abstract}
In some quantum many-body systems, the Hilbert space breaks up into a large ergodic sector and a much smaller scar subspace. It has been suggested
[K. Pakrouski et al., Phys. Rev. Lett. 125, 230602 (2020)]
that the two sectors may be distinguished by their transformation properties under a large group whose rank grows with the system size (it is not a symmetry of the Hamiltonian). The quantum many-body scars are invariant under this group, while all other states are not. Here we apply this idea to lattice systems containing $M$ Majorana fermions per site. The Hilbert space for $N$ sites may be decomposed under the action of the $\SO(N)\times \SO(M)$ group, and the scars are the $\SO(N)$ singlets. For any even $M$ there are two families of scars. One of them, which we call the $\eta$ states, is symmetric under the group O$(N)$. The other, the $\zeta$ states, has the SO$(N)$ invariance. For $M=4$, where our construction reduces to spin-$1/2$ fermions on a lattice with local interactions, the former family are the $N+1$ $\eta$-pairing states, while the latter are the $N+1$ states of maximum spin. We generalize this construction to $M>4$. For $M=6$ we exhibit explicit formulae for the scar states and use them to calculate the bipartite entanglement entropy analytically. For large $N$, it grows logarithmically with the sub-system size. We present a general argument that any group-invariant scars should have the entanglement entropy that is parametrically smaller than that of typical states. The energies of the scars we find are not equidistant in general but can be made so by choosing Hamiltonian parameters. For $M>6$ we find that with local Hamiltonians the scars typically have certain degeneracies. The scar spectrum can be made ergodic by adding a non-local interaction term. We derive the dimension of each scar family and show the scars could have a large contribution to the density of states for small $N$.
\end{abstract}

\maketitle

%{
%  \hypersetup{linkcolor=black}
 % \tableofcontents
%}

\section{Introduction}

The past few years have seen growing interest in ``quantum many-body scars," the term that was coined in \cite{Turner_2018}.
The essential phenomenon is that there are many-body systems where the Hilbert space breaks up into the bulk of states that satisfy the 
Eigenstate Thermalization Hypothesis (ETH), and a much smaller scar subspace that does not.
Specific constructions of such states have been found in a variety of models \cite{PhysRevA.86.041601,Shiraishi2017ScarsConstruction,Turner_2018,Moudgalya:2018,AbaninScarsSU2Dynamics,Khemani:2019vor,Sala_2020,Prem:2018,Schecter:2019oej,SciPostPhys.3.6.043,IadecolaHubbardAlmostNUPRL2019,Shibata:2020yek,michailidis2020stabilizing,2020MarkMotrEtaPairHub,PRLPapicClockModels,VedikaScarsVsIntegr,Pal2020ScarsFromFrustration,mark2020unified,iadecola2020quantum,moudgalya2020etapairing,PhysRevB.101.220305,pakrouski2020GroupInvariantScars,PhysRevLett.126.120604,PhysRevResearch.2.043305,Nielsen2020ChiralScars,Hsieh2020PXP2D,Regnault2020MPStoFindSc,FloquetPXPScars,PapicWeaklyBrokenAlgebra,kuno2021multiple,banerjee2020quantum,Pakrouski:2021jon,chaoticDickeAllScarred2021,maskara2021DrivenScars,langlett2021rainbow,2021arXiv210807817R,2021arXiv211011448T,japaneseGeneralizedEtaPairingSep2021,Schindler:2021lma,
Barnes:2022qli,Moudgalya:2022gtj,Moudgalya:2022nll,Nakagawa:2022jsg,PhysRevB.106.235147,Dodelson:2022eiz,Caputa:2022zsr,Liska:2022vrd,BucaNature2019}. For the recent reviews of the different approaches to scars, see
\cite{Serbyn:2020wys,Moudgalya:2021xlu,Papic2022,Chandran:2022jtd}. A related phenomenon of non-stationary dynamics has also been identified in open systems \cite{BucaNature2019,HadisehMultistability2022}.

Remarkably, the quantum many-body scars appear in the commonly used models of condensed matter physics, such as the (deformed) Fermi-Hubbard and t-J-U models on a lattice with $N$ sites. 
Such models
contain two species of complex fermions on each site, $c_{j\uparrow}$ and $c_{j\downarrow}$. In addition to the rotational $\SU(2)$ symmetry, they possesses a (broken) pseudospin $\widetilde{\SU}(2)$ symmetry. The $\eta$-pairing states \cite{etaPairingYang89,yang1990so}
form a multiplet of pseudospin $N/2$, and their role as scars was pointed out and studied in \cite{2020MarkMotrEtaPairHub,moudgalya2020etapairing,pakrouski2020GroupInvariantScars,Pakrouski:2021jon}. Another important family are the $\zeta$-states that carry the maximum spin $N/2$; they can be regarded as scars if the $\SU(2)$ rotational symmetry is broken \cite{pakrouski2020GroupInvariantScars,Pakrouski:2021jon}.

There has also been important progress on generalizing the $\eta$-pairing states to systems with more than two complex fermions per lattice site \cite{japaneseGeneralizedEtaPairingSep2021,Nakagawa:2022jsg} (for earlier work, see also \cite{ZhaiGeneralizedEtaPairing2005}).

In this paper we present a systematic method for constructing multiflavor fermionic systems with weak ergodicity breaking (many-body scars) that leads to generalizing the $\eta$ and $\zeta$ states. It relies on the idea that the scar subspace is invariant under a ``large group" whose rank is of the order of the number of lattice sites $N$ \cite{pakrouski2020GroupInvariantScars,Pakrouski:2021jon}. Indeed, the $\eta$-pairing scars in the spin-$1/2$ Hubbard model have been shown to be invariant under $\SO(N)$ which acts on the lattice site index
\cite{pakrouski2020GroupInvariantScars}, as well as under an even bigger symplectic group $\tSp (N)$ \cite{Pakrouski:2021jon}.

We consider lattice systems of arbitrary dimension with an even number $M$ of Majorana fermions per lattice site and rely on the method proposed in \cite{pakrouski2020GroupInvariantScars}: the many-body scars span a subspace of the Hilbert space invariant under a large group $G$ that is
{\it not} a symmetry of the system. The Hamiltonian is chosen to be of the form $H=H_0+\sum_a \CO_a T_a$, where $H_0$ is a term governing the dynamics of the scar subspace,
$T_a$ are generators of the group $G$, which therefore annihilate the scars, and $\CO_a$ are operators chosen so that the Hamiltonian is Hermitian. Here, we apply this construction using the group $G=\SO(N)$ and $H_0$ given by the (generalized) Hubbard model, which commutes with the quadratic Casimir operator of $\SO(N)$. We also exhibit other operators that can be used as building blocks of scarred Hamiltonians in this Hilbert space, thus providing a toolbox for future studies.

Our system is equivalent to $M/2$ complex fermions per site, and for $M=4$ we reproduce the previous results singling out $N+1$ $\eta$-states and $N+1$
$\zeta$-states as the scars. For $M=6$ we present concise formulae for all the $\SO(N)$ singlets, which come in two families generalizing the $\eta$ and $\zeta$ states.  The expressions for the generalized $\eta$-states are the same as those found in \cite{Nakagawa:2022jsg}, where it was also understood that they are eigenstates of the $\SU(3)$ invariant Hubbard interaction.
We calculate the bipartite entanglement entropy for $M=6$ analytically. For small $N$ we confirm numerically that the invariant states we consider possess all the characteristic properties of many-body scars. In particular their entanglement entropy is found to be much smaller than that of the nearby thermal states. For large $N$ we show analytically that the entanglement entropies of scars grows logarithmically with the region size for any $M$.

Practically all the many-body scars known in literature are characterized by the lower entanglement entropy compared to generic states in the same energy range. This feature is even sometimes included into the definition of scars. However, the underlying mechanism behind this has remained unclear. Generalising our results for the Majorana scars we present in Sec. \ref{sec:entrBound} a general argument that any group-invariant scars in any system built according to the group-invariant formalism \cite{pakrouski2020GroupInvariantScars} must have a parametrically lower entanglement and conjecture that the entropy of other, not group-invariant scars is reduced for a similar reason.

For $M>4$, the scars within the $\eta$ and $\zeta$ families are not in general equidistant in energy even if the Hamiltonian is restricted to local terms only. We discuss the conditions under which the "revivals" can nevertheless be observed. Because of the large number of scars (this is also the case in the model of \cite{BiaoLian}), their presence is noticeable in the density of states for small $N$ which has implications for their experimental detection. Furthermore, the scar spectrum becomes ergodic if the $H_0$ part \cite{pakrouski2020GroupInvariantScars} of the Hamiltonian is chosen to be non-integrable. These features of many-body scar spectrum are reported here for the first time to our knowledge.

For $M>6$ the $\SU(M/2)$ invariant Hubbard interaction no longer works simply. Therefore, we replace it by another local interaction under which all the $\SO(N)$ singlets are eigenstates. A novel feature we find for the scar states with $M>6$ is the 
presence of degeneracies which appear to be protected from any local interactions, but can be broken by non-local ones.

\section{Deformed Hubbard model and scars \label{sec:deformedHubbardM4}}

In this section, we review the $\SO(4)$ symmetry of the Hubbard model, using both Dirac and Majorana fermions,  and discuss its relation with the scar states in some deformed Hubbard models.
For simplicity of the discussion, we consider here the model on a 1D lattice of $N$ sites but the results hold \cite{Pakrouski:2021jon} in any dimension. The standard Hubbard Hamiltonian is 
the sum of three terms -- the hopping, the on-site repulsion, and the chemical potential:
\begin{align}\label{HubbardHam}
&T=-t\sum_{j=1}^{N-1} \sum_{\sigma\in\{\uparrow,\downarrow\}} \left(c^{\dagger}_{j\sigma}c_{j+1,\sigma}+ c^{\dagger}_{j+1,\sigma}c_{j\sigma} \right)\ ,  \nonumber\\
&V=U\sum_{j=1}^N n_{j\uparrow} n_{j\downarrow}=U\sum_{j=1}^N c^\dagger_{j\uparrow}c_{j\uparrow}c^\dagger_{j\downarrow}c_{j\downarrow}\ , \qquad \nonumber\\
& \mu = - \sum_{j=1}^N \left (\mu_{\downarrow} c^\dagger_{j\downarrow}c_{j\downarrow} +  \mu_{\uparrow} c^\dagger_{j\uparrow}c_{j\uparrow} \right )\ .
\end{align}
Here $t$ is a real hopping parameter, $U>0$ is the on-site interaction strength,
 and $c_{i\sigma}, c^\dagger_{i\sigma}$ are the fermionic ladder operators satisfying the anticommutation relations
\begin{gather}
    \left\{c_{i \sigma}, c^\dagger_{j\sigma'} \right\} = \delta_{ij} \delta_{\sigma\sigma'}\ .
\end{gather}
The magnetic field is $(\mu_{\uparrow}- \mu_{\downarrow})/2$ while the standard chemical potential is $(\mu_{\uparrow}+ \mu_{\downarrow})/2$.
We find it convenient to perform a site dependent phase rotation, $c_{j\sigma}\to e^{-i j\frac{\pi}{2}}c_{j\sigma}$, upon which the hopping term acquires an imaginary coefficient.
\begin{equation}
T' = it\sum_{j=1}^{N-1} \sum_{\sigma\in\{\uparrow,\downarrow\}} \left(c^{\dagger}_{j\sigma}c_{j+1,\sigma}- c^{\dagger}_{j+1,\sigma}c_{j\sigma} \right)\ , \label{eq:twistHam}
\end{equation}
In these variables, the Hubbard Hamiltonian is $H_{Hub}=T'+V+\mu$. This transformation could be performed on any dimensional bipartite lattice \cite{Pakrouski:2021jon}. We shall not restrict ourselves only to bipartite lattices and consider the Hamiltonian \eqref{eq:twistHam} on any lattice.

The time reversal symmetry in this case is defined as follows
\begin{gather}
\text{T}\, c_{i \uparrow} \text{T}^{-1} = (-1)^i c_{i \downarrow}, \quad \text{T}\, c_{i \downarrow} \text{T}^{-1} = (-1)^{i+1} c_{i \uparrow}
\end{gather}

For the vanishing magnetic field, $\mu_{\uparrow}= \mu_{\downarrow}$, the Hamiltonian $H_{Hub}$ has both time reversal symmetry and  the spin $\SU(2)$ symmetry, which acts on the spin index $\sigma$. The  generators of  $\SU(2)$ are 
\begin{align}
&\zeta_3=\frac{1}{2}\sum_j\left(c^\dagger_{j\uparrow}c_{j\uparrow}-c^\dagger_{j\downarrow}c_{j\downarrow}\right), \\ \notag\,\,\,\,\, &\zeta^+=\sum_j c^\dagger_{j\uparrow}c_{j\downarrow}, \,\,\,\,\,\zeta^-=\sum_j c^\dagger_{j\downarrow}c_{j\uparrow}\ .
\end{align}
When $\mu_{\uparrow}= \mu_{\downarrow}=\frac{U}{2}$, the $\SU(2)$ symmetry is enhanced to $\SO(4)= \SU(2)\times \widetilde{\SU}(2)/\mathbb Z_2$, where the pseudospin group  $\widetilde{\SU}(2)$ is generated by \cite{etaPairingYang89,yang1990so,ZhangHubbardSO41991}
\begin{align}\label{etadef}
&\eta_3=\frac{1}{2}\sum_j\sum_\sigma c^\dagger_{j\sigma}c_{j\sigma}-\frac{N}{2}, \\ \notag \,\,\,\,\, &\eta^+=\sum_j c^\dagger_{j\uparrow}c^\dagger_{j\downarrow}, \,\,\,\,\,\eta^-=\sum_j c_{j\downarrow}c_{j\uparrow}\ ,
\end{align}
and the $\mathbb Z_2$ factor is realized by the Shiba transformation \cite{Shiba1972TransformOriginal, hubbard1DbookShiba, Pakrouski:2021jon}. The $\SO(4)$ symmetry becomes manifest if we use a Majorana representation of the Dirac spinors $c_{j\sigma}$. Namely, we define
\begin{align}
c_{j\uparrow}=\frac{\psi_{j}^{1}-i \psi_{j}^{2}}{\sqrt{2}}\ , \,\,\,\,\, c_{j\downarrow}= \frac{\psi_{j}^{3}-i \psi_{j}^{4}}{\sqrt{2}},
\end{align}
where $\psi_j^A, A=1,2,3,4$ are four Majorana fermions on site $j$. Then $H_{Hub}$ at $\mu_\uparrow=\mu_\downarrow=\frac{U}{2}$ admits a manifestly $\SO(4)$ invariant form
\small
\begin{align}\label{SO4Hub}
H_{Hub}=it\sum_{j}\sum_{A=1}^4 \psi^A_j\psi^A_{j+1}-U\sum_j \psi_j^1\,\psi_j^2\,\psi_j^3\,\psi_j^4-\frac{UN}{4}
\end{align}
\normalsize
where the hopping terms $\sum_A \psi^A_j\psi^A_{j+1}$ can be identified as special cases of the (antihermitian) $\SO(N)$ generators  $T_{jk}\equiv \sum_{A}\psi^A_j\psi^A_k, 1\le j<k\le N$. The $\SO(N)$ group acts on the site indices of Majorana fermions, i.e. $\psi_j^A\to R_{ij} \psi_j^A$, where $R_{ij}$ is a special orthogonal matrix. The $\SO(N)$ singlet subspace consists of $\eta$ states 
 \begin{align}\label{etastate}
 |m^\eta\rangle \equiv  \frac{\left(\eta^+\right)^m}{\sqrt{\frac{N!m!}{(N-m)!}}}|0^\eta\rangle, \,\,\,\,\, m=0,1,2,\cdots,N
 \end{align}
where the $\eta$-vacuum $|0^\eta\rangle$ is the same as the empty vacuum $|0\rangle$, and the $\zeta$ states 
 \begin{align}\label{zetastate}
 |m^\zeta\rangle \equiv \frac{\left(\zeta^+\right)^m}{\sqrt{\frac{N!m!}{(N-m)!}}}|0^\zeta\rangle,  \,\,\,\,\, m=0,1,2,\cdots,N
 \end{align}
 where the $\zeta$-vacuum is $|0^\zeta\rangle\equiv c^\dagger_{1\downarrow}\cdots c^\dagger_{N\downarrow}|0\rangle$. The $\eta$ states span an $(N+1)$ dimensional representation of the pseudospin group $\widetilde {\SU}(2)$ and the $\zeta$ states furnish a spin $\frac{N}{2}$ representation of the  spin group $\SU(2)$. The $\zeta$ states have fixed  fermion number $N$ and have eigenenergies $-(\mu_\uparrow-\mu_\downarrow)m-\mu_\downarrow N$ with respect to the Hubbard Hamiltonian $H_{Hub}$.
The $\eta$ states are also energy eigenstates of the Hubbard Hamiltonian, i.e. $H_{Hub}|m^\eta\rangle = m(U-\mu_\downarrow - \mu_\uparrow)  |m^\eta\rangle$.
In \cite{Pakrouski:2021jon}, the authors constructed the Hubbard model deformed by quartic $\rm OT$ terms that break both the spin and pseudospin symmetries. Then the $\SO(N)$ invariant $\eta$ and $\zeta$ states remain eigenstates, and they have all the typical properties of scar states. In the next section we will consider a different set of symmetry breaking deformations which also 
keep the $\eta$ and $\zeta$ states as scars. We will also extend the construction from $4$ Majorana fermions per site to a higher even number $M$.

\section{Multi-flavor Majorana fermions on a lattice}

Generalizing the Majorana description of the Hubbard model, we consider  a lattice of $N$ sites, hosting an even number $M$ flavors of Majorana fermions $\psi^A_j, A=1,2,\cdots, M$ on each site. Their anti-commutation relations 
\begin{equation}
\label{eq:antiCommEq}
\{\psi^A_i, \psi^B_j\}=\delta^{AB}\delta_{ij}
\end{equation}
are invariant under the action of $\SO(N) \times \SO(M)$ group.  We can build (antihermitian) generators of $\SO(N)$ and $\SO(M)$ out of these fermions
\begin{gather}
\label{eq:SONgenerators}
T_{ij}=\frac{1}{2}\sum_{A=1}^{M} [\psi^A_i, \psi^A_j], \quad 
J^{A B}=\frac{1}{2}\sum_{j=1}^N [\psi^A_j, \psi^B_j]\ .
\end{gather} 
Their commutation relations are given by 
\begin{align}\label{JJ}
[T_{ij}, T_{kl}]&=\delta_{jk} T_{il}-\delta_{ik} T_{jl}-\delta_{jl} T_{ik}+\delta_{il} T_{jk}\\
[J^{AB}, J^{CD}]&=\delta^{BC} J^{AD}\!-\!\delta^{AC} J^{BD}\!-\!\delta^{BD} J^{AC}\!+\!\delta^{AD} J^{BC}\nonumber\\
[T_{ij}, J^{AB}]&=0\ .\nonumber
\end{align}

Grouping the Majorana fermions into complex combinations provides a convenient way to construct states in the Hilbert space.  On each site $j$, define $\alpha=1,\ldots,M/2$ flavors of complex fermions
\begin{align}
c_{j\alpha}=\frac{\psi_{j}^{2\alpha-1}-i \psi_{j}^{2\alpha}}{\sqrt{2}}\ , \,\,\,\,\, c_{j\alpha}^{\dagger}= \frac{\psi_{j}^{2\alpha-1}+i \psi_{j}^{2\alpha}}{\sqrt{2}}
\end{align}
which satisfy the standard anticommutation relations $\{c_{i\alpha}, c_{j\beta}^\dagger\}=\delta_{\alpha \beta}\delta_{ij}$. Following  the general recipe described in appendix \ref{SOreview}, we construct a basis of $\so(M)$ in terms of these complex fermions that makes its root decomposition structure manifest:
\begin{align}\label{rootcc}
&\text{Cartan generators}:\,\,\, h_\alpha=\sum_{j} c^{\dagger}_{j\alpha} c_{j\alpha}-\frac{N}{2}\\
&\text{Raising operators}:\,\, \, 
 \zeta^\dagger_{\beta\gamma}= \sum_j c^{\dagger}_{j\beta}c_{j\gamma}, \quad 
 \eta^\dagger_{\beta\gamma}= \sum_j c^{\dagger}_{j\beta} c_{j\gamma}^{\dagger}\nonumber\\
&\text{Lowering operators}:\,\, \, 
\zeta_{\beta\gamma}=\sum_j c^{\dagger}_{j\gamma} c_{j\beta}, \quad 
\eta_{\beta\gamma}=\sum_j c_{j\gamma} c_{j\beta}\nonumber
\end{align}
where $1\le \beta<\gamma\le M/2$. The Cartan generators $h_\alpha$ count fermion numbers of each flavor $\alpha$ (up to the constant shift $-\frac{N}{2}$). And the raising and lowering operators hop the fermions in a particular direction (in the flavor space). They correspond to the positive and negative roots of $\so(M)$ (defined in appendix \ref{SOreview}) respectively.

The full Hilbert space $\CH$ has dimension $2^{MN/2}$ and forms a spinor representation of the group $\SO(MN)$ which acts on the collective index $I=(A;i)$. The decomposition of $\CH$ under the subgroup
$\SO(N)\times \SO(M)$ was studied in \cite{Klebanov:2018nfp,Gaitan:2020zbm}. The quadratic Casimir operators of $\SO(N)$ and $\SO(M)$, defined by eq. (\ref{Casdefo}), satisfy the sum rule \cite{Klebanov:2018nfp}
\begin{align}
\CC_{2}^{\SO(N)}+ \CC_{2}^{\SO(M)} =  \frac{1} {8} MN (M+N- 2)\ .
\label{casimirsum}
\end{align}
In particular, the $\SO(N)$ singlets have $\SO(M)$ Casimir $\frac{1} {8} MN (M+N- 2)$, which according to eq. (\ref{specas}) implies that these singlet states furnish representations of highest weight or their direct sums 
\begin{align}
\label{eq:lambdaHiWeight}
\lambda_{N/2}^\pm\equiv \frac{1}{2}(\underbrace{N,N,\cdots,\pm N}_{M/2}),
\end{align} 
where the right hand side stands for a rectangular Young tableaux with $N$ columns of height $M/2$. When $N$ is even, it was found in \cite{Gaitan:2020zbm} that the $\text{SO}(N)$ singlets furnish the reducible representation $\lambda_{N/2}^+\oplus \lambda_{N/2}^-$ by using a character method. We will present an elementary way to show that this structure holds for both odd and even $N$ in section \ref{Singsec}, where we study in more detail the structure of these singlets.

\subsection{Ergodic Hamiltonians that support singlet states as many-body scars}\label{erere}

Following the recipe proposed in \cite{pakrouski2020GroupInvariantScars}, we construct (local) Hamiltonians of the form $H=H_0+\sum\CO_{ij} T_{ij}$. 
The first term $H_0$ is designed to leave the space of $\SO(N)$ singlets (denoted by $\mathbb{S}$) invariant. 
The second term, which will be referred to as the $\rm OT$ term henceforth, should be hermitian and break some symmetries of $H_0$, by choosing $\CO_{ij}$ properly. The Hamiltonians we discuss below should have as few symmetries as possible such that they are guaranteed to produce ergodic bulk spectra and such that the many-body scars are not fully occupying an isolated sector of a symmetry. Nevertheless the Hamiltonians we consider always preserve the fermion number parity symmetry
 \begin{gather}
 \label{eq:fermParity}
P_f = (-1)^Q, \quad Q=\sum_{j=1}^N\sum_{\alpha=1}^{M/2} n_{j\alpha}, \quad \left[P_f, H\right] = 0,
 \end{gather}
 where $n_{j\alpha}=c^\dagger_{j\alpha} c_{j\alpha}$ is the fermion number operator of flavor $\alpha$ at site $j$.
 All the numerical computations presented in Sec.\ref{sec:numerics} are performed separately in each of the two sectors.

The Hubbard Hamiltonian in eq. (\ref{SO4Hub}) admits a straightforward $\SO(M)$-invariant generalization to the $M$-flavor Majorana model on any lattice, namely 
$H_M\equiv T_M+H_\mu+V_M$, where 
\begin{align}
\label{eq:allFlavourUakaVsubM}
&T_M=it\sum_{\langle j,k\rangle} T_{jk}=it \sum_{\langle j,k\rangle}\sum_{A=1}^M\psi^A_j\psi^A_k\ , \qquad \nonumber \\
&H_\mu=-\sum_{\alpha} \mu_\alpha \left(h_\alpha+\frac{N}{2}\right)=-\sum_\alpha\, \mu_\alpha \sum_{j} n_{j\alpha}\ , \qquad\nonumber\\
&V_M=U i^{\frac{M}{2}}\sum_{j} \psi^1_j\psi^2_j\cdots\psi_j^M-2^{-\frac{M}{2}}NU\ .
\end{align}
The hopping term $T_M$ is summed over nearest neighbors. It can be thought as a special $\rm OT$ term with $\CO_{ij}\equiv it$ when $(i,j)$ are nearest neighbors and $\CO_{ij}=0$ otherwise. The $V_M$ term leaves  $\mathbb S$ invariant because the states in $\mathbb S$ have a fixed $\SO(M)$ Casimir $\frac{MN}{8}(M+N-2)$ and $V_M$ commutes with the $\SO(M)$ generators. 
Then the sum rule (\ref{casimirsum}) implies that $V_M\mathbb S\subset \mathbb S$. Furthermore,
\begin{align}
\label{eq:hidden}
[ H_M, \CC_{2}^{\SO(N)}] = 0\ ,
\end{align} 
which is a special case of the criterion proposed in eq. (1) of \cite{pakrouski2020GroupInvariantScars}. This equation means that the Hubbard model, as well as its higher $M$ 
generalizations (\ref{eq:allFlavourUakaVsubM}),
factorizes the Hilbert space into the representations of $\SO(N)$.
$\SO(N)$ symmetry is broken in a way that only creates energy splittings within the representations but does not mix different $\SO(N)$ representations. 

In the complex fermion basis, 
\begin{align}\label{H0c}
&T_M= i t \sum_{\langle j,k\rangle}\sum_{\alpha}\left(c^\dagger_{j\alpha}c_{k\alpha}\!-\!c^\dagger_{k\alpha}c_{j\alpha}\right)\ , \nonumber \\
& V_M= U\sum_{j}\prod_\alpha\left(\frac{1}{2}\!-\!n_{j\alpha}\right) -2^{-\frac{M}{2}}NU\ .
\end{align}
The constant shift insures that $V_M$ vanishes when all $n_{j\alpha}$ are equal to zero.
For $M=4$, $T_4+V_4+H_\mu$ is equivalent to the standard Hubbard Hamiltonian with chemical potentials $\mu_\uparrow=\mu_1+\frac{U}{2}$, and 
$\mu_\downarrow=\mu_2+\frac{U}{2}$.
For general $M$, since $n_{j\alpha}$ takes value in $\{0,1\}$, we have $\frac{1}{2}-n_{j\alpha}=\frac{1}{2}(-1)^{n_{j\alpha}}$ and hence $V_M$ can be simplified as
\begin{align}
V_M=U\sum_j  \frac{(-1)^{n_j}-1}{2^{\frac{M}{2}}}\ , \qquad  n_j=\sum_{\alpha=1}^{M/2} n_{j\alpha}\ ,
\end{align}
where $n_j$ is the total fermion number at site $j$. 

The  chemical potentials in $H_\mu$
break the $\SO(M)$ symmetry of $H_M$. 
In general, to break $\SO(M)$, we could start with the more general
$\sum_{AB} i \mu_{AB} J^{AB}$, where $\mu_{AB}$ is a real anti-symmetric matrix, and then
redefine the Majorana fermions to get $H_\mu$ (up to a constant shift).
While $H_\mu$ breaks the $\SO(M)$ symmetry, the $\SO(M)$ Casimir $\CC_2^{\SO(M)}$ is still a conserved charge, 
since all $h_\alpha$ in $H_\mu$ commute with $\CC_2^{\SO(M)}$.

In the Hubbard model \eqref{HubbardHam}, generic chemical potentials $\mu_\uparrow$ and $\mu_\downarrow$ break the $\SO(4)$ symmetry. For the Hamiltonian $H_M$, the Casimir operators of $\SU(2)$ and $\widetilde{\SU}(2)$ (consider $M=4$) are conserved charges because the symmetry breaking term is a linear combination of $\eta_3$ and $\zeta_3$.
The $\rm OT$ term can be used to break the conservation of $\CC_2^{\SO(M)}$. For instance, the following sextic interacting term does this job
\begin{gather}
\label{eq:TOT}
\widetilde H_{\rm int}= \sum_{\langle j,k\rangle} T_{jk} \left(i\sum_{A< B}r_{AB}\, \psi^A_j\psi^B_j\right) T_{jk}\ ,
\end{gather}
where $r_{AB}$ are randomly chosen real numbers. This form of local interacting Hamiltonian is similar to that adopted by Shiraishi and Mori \cite{Shiraishi2017ScarsConstruction}, 
$\sum_j P_j h_j P_j$, where $P_j$ is a set of local projection operators that satisfy $P_j^2=P_j$. While our $\SO(N)$ generators $T_{jk}$ are not projectors and do not commute, our singlet conditions on the scar states, $T_{jk}|\phi\rangle = 0$ are analogous to the conditions $P_j|\phi\rangle = 0$ imposed in \cite{Shiraishi2017ScarsConstruction}.
In the recent papers \cite{Moudgalya:2022gtj,Moudgalya:2022nll}, some parallels were drawn between the approach identifying scars as the sector invariant under a large group \cite{pakrouski2020GroupInvariantScars,Pakrouski:2021jon}
and the Shiraishi-Mori construction. 

Adding up $H_M$ and $\widetilde H_{\rm int}$, we get the following Hamiltonian
\begin{align}\label{FullHam}
H=&\, - \sum_\alpha \mu_\alpha \sum_j n_{j\alpha}+U i^{\frac{M}{2}} \sum_{j}\psi^1_j\psi^2_j\cdots\psi_j^M -2^{-\frac{M}{2}}NU\nonumber \\
&+it \sum_{\langle j,k\rangle}T_{jk}+\sum_{\langle j,k\rangle} T_{jk} \left(i\sum_{A< B}r_{AB}\, \psi^A_j\psi^B_j\right) T_{jk}\ .
\end{align}
In terms of the general scheme \cite{pakrouski2020GroupInvariantScars}, $H_0=H_M$ and $H_1= \widetilde H_{\rm int}$.
Later, in Sec. \ref{sec:NumM4} we confirm numerically that the Hamiltonian \eqref{FullHam} indeed exhibits weak ergodicity breaking.

Alternatively, instead of the sextic interaction $\rm OT$ term (\ref{eq:TOT}), we may consider the quartic one
\begin{gather}
\label{eq:OT}
	H_{\rm OT} = i\sum_{\langle j,k\rangle} \CO_{jk} T_{jk},
\end{gather}
with $\CO_{jk}$ being hermitian, quadratic in Majorana operators and satisfying $\left[\CO_{jk}, T_{jk}\right] = 0$. It is easy to check  such $H_{\rm OT}$ is  hermitian and annihilates the singlets. There is a simple way to construct $\CO_{jk}$ by noting that $T_{jk}$ can be regarded as a generator of a $U(1)$ $Q_T$ symmetry. This gives the following charges
\begin{gather*}
	Q_T\left(\psi_{i}^A\right) = 0, \quad Q_T\left(d_\pm^{jk,A}\right) = \pm 1, \quad d_\pm^{jk,A} = \psi^A_j \pm i \psi^A_k \ .
\end{gather*}
An operator $\CO_{jk}$ that has zero charge $Q_T$ then commutes with the hopping operator $T_{jk}$. For instance, we can consider the following term, that is biliniear in $d_\pm^{jk,A}$
\begin{gather}
\label{hjk}
 H_{\rm OT} = i\sum_{\langle j,k\rangle}\left(\sum_{AB} R_{AB}d_+^{jk,A} d_-^{jk,B}\right) T_{jk}, \quad R^*_{AB} = R_{BA}.
\end{gather}
In Sec. \ref{sec:NumMlt4} we provide numerical evidence that the Hamiltonian that includes \eqref{hjk} also supports many-body scars.

Another natural generalization of the Hubbard model would be on-site density-density interaction between different flavors, which can be described by the following potential \cite{PhysRevLett.92.170403}   
\begin{align}\label{SUNHubbard}
\widetilde V_M=\widetilde U \sum_j \sum_{\alpha<\beta} n_{j\alpha} n_{j\beta}
\end{align}
This term does not have $\SO(M)$ symmetry for $M\ge 6$ which was also noticed in \cite{Nakagawa:2022jsg} although $\widetilde V_M$ still keeps some of the scar states considered in \cite{Nakagawa:2022jsg} invariant.

\subsection{Controlling the position of scars in the spectrum \label{sec:controlScarPos}}

One of the possible strategies \cite{Pakrouski:2021jon} to control the position of scars in the spectrum relies on the addition of a term that annihilates all the scars but acts positive-definitely on all other states. Because scars 
are  $\SO(N)$-invariant in our case the most obvious choice for such a term is the quadratic Casimir operator of the $\SO(N)$ group
\begin{align}
\label{eq:CasSON}
\CC_2^{\SO(N)} = -\frac{1}{2}\sum_{i=1}^N \sum_{j=1}^N T^2_{ij} 
\end{align} 
where $T_{ij}$ are $\SO(N)$ generators given in eq. \eqref{eq:SONgenerators}. The interaction in eq. \eqref{eq:CasSON} is however highly non-local in real space.

We find numerically that for the accessible system sizes also the following local interaction can be used where the summation is only over the nearest neighbours
\begin{align}
\label{eq:HUTsq}
H_{T^2} =  - \sum_{\langle ij\rangle} T^2_{ij}.
\end{align} 
It is non-negative definite, and because of the presence of $\SO(N)$ generators, we have  
\begin{align}
H_{T^2}\ket{s} = 0,\quad  \langle ns | H_{T^2}| ns \rangle > 0,
\end{align}
where $\ket{s}$ is an $\SO(N)$ singlet state and $\ket{ns}$ is a non-singlet state. 
Therefore, it can be used to change the position of the scars in energy with respect to all other states without changing the relative position (and the revivals period) of scars themselves. In particular, using eq. \eqref{eq:HUTsq} one can achieve that the low-energy part of the spectrum is comprised of many-body scars only as shown in Fig. \ref{fig:scarsDOS}.

 Another strategy for scar spectrum design is also inspired by the single-band case \cite{Pakrouski:2021jon} where the scars furnished the highest-spin representation of $\SU(2)$ and would therefore be the most susceptible states to the magnetic field which in turn allows to make a scar the ground state.

Similarly, the scar subspace in the present case corresponds to the highest-weight representation of $SO(M)$ (see Sec. \ref{sec:soMstructOfSingl}) and includes the states most susceptible to the chemical potential \eqref{eq:allFlavourUakaVsubM} $H_\mu$ part of the Hamiltonian. Therefore by a suitable choice of the strong chemical potential (playing here a role of a generalized magnetic field) one can always achieve that a scar state becomes the ground state.

\section{$\SO(N)$ singlets as scars}\label{Singsec}

In this section we discuss properties of the $\SO(N)$ singlets for arbitrary $M$. For $M=6$, i.e. $6$ Majorana fermion flavors per site, we write down the wavefunction of every $\SO(N)$ singlet.
Similar explicit formulae for the $\eta$ states have been obtained in the $\SU(3)$ Hubbard model \cite{Nakagawa:2022jsg}.
Furthermore, we show analytically that 
both the $\eta$ states and the $\zeta$ states
have a sub-volume law for entanglement entropy in the thermodynamic limit.

\subsection{The $\SO(M)$ representation structure of $\SO(N)$ singlets \label{sec:soMstructOfSingl}}

From the Casimir relation (\ref{casimirsum}), we know that the possible $\SO(M)$ representation structure of the $\SO(N)$ singlets is a direct sum of the maximal-weight representations from eq. \eqref{eq:lambdaHiWeight}:
\begin{align}\label{jo;lj}
\mathbb{S} =\underbrace{\lambda^+_{N/2} \oplus\cdots \oplus  \lambda^+_{N/2}}_{n_+}\oplus \underbrace{\lambda^-_{N/2} \oplus\cdots \oplus  \lambda^-_{N/2}}_{n_-}
\end{align}
where $n_\pm$ are multiplicities of each $\lambda^\pm_{N/2}$ representation.
Given a highest weight representation $R$, its lowest-weight vector is given by minus of the highest-weight vector of the dual representation $R^*$. When $n=M/2$ is even, all highest-weight representations of $\SO(2n)$ are self-dual, and when $n$ is odd, the dual representation is given by flipping the sign of the last entry of the highest-weight vector. More explicitly, the lowest-weight vector corresponding to $(\lambda_1,\lambda_2,\cdots, \lambda_n)$ is $(-\lambda_1,-\lambda_2,\cdots, (-)^{n+1}\lambda_n)$.
For the two irreducible representations in eq. (\ref{jo;lj}), the corresponding lowest-weight vectors are $-\frac{1}{2}(N,N,\cdots,\pm N)$.

This implies that the lowest-weight states have 0 occupations for $\alpha=1,2,\cdots, M/2-1$ and have either 0 or $N$ occupations for flavor $\alpha=M/2$. These conditions completely fix the possible lowest-weight states:
\begin{align}
|0\rangle, \,\,\,\,\, |0^\zeta\rangle\equiv c_{1,M/2}^{\dagger}c_{2, M/2}^{\dagger}\cdots c_{N, M/2}^{\dagger}|0\rangle
\end{align}
where $|0^\zeta\rangle$ can be thought as the multi-flavor generalization of the $\zeta$-vacuum. They are both manifestly $\SO(N)$ invariant. The state $|0\rangle$ is annihilated by $\eta_{\alpha\beta}$, $\zeta_{\alpha\beta}$ and $\zeta^\dagger_{\alpha\beta}$. 
The state $|0^\zeta\rangle$ is  annihilated by all $\eta_{\alpha\beta}$ and $\zeta_{\alpha\beta}$, and  $\zeta^\dagger_{\alpha\beta}$ with $\beta<M/2$, because they  contain at least one annihilation operator of flavor $1,2, \cdots, M/2-1$. It is also annihilated by $\eta^\dagger_{\alpha,M/2}$ since it is fully filled at flavor $M/2$.

Denote the irreducible representation containing $|0\rangle$ by $\CH_\eta$, which corresponds to the highest-weight vector $\lambda_\eta=\left(\frac{N}{2},\frac{N}{2},\cdots,(-1)^{M/2}\frac{N}{2} \right)$, and denote the irreducible representation containing $|0^\zeta\rangle$ by $\CH_\zeta$, which corresponds to the highest-weight vector $\lambda_\zeta=\left(\frac{N}{2},\frac{N}{2},\cdots,(-1)^{M/2+1}\frac{N}{2} \right)$. Then the singlet subspace $\mathbb S$ is a direct sum of $\CH_\eta\oplus \CH_\zeta$, i.e. $n_+=n_-=1$ in eq. (\ref{jo;lj}).  States in $\CH_\eta$ are obtained by acting with all 
\begin{align}
\eta^\dagger_{\alpha\beta}=\sum_j c_{j\alpha}^\dagger c_{j\beta}^\dagger
\end{align}
repeatedly on the empty vacuum $|0\rangle$, and hence  are generalizations of $\eta$ states in the Hubbard model. Similarly, $\CH_\zeta$ is built with $\zeta^\dagger_{\beta,M/2}=\sum_j c^\dagger_{j\beta}c_{j,M/2}$ ($1\le \beta\le M/2-1$) and $\eta^\dagger_{\alpha\beta}$ ( $1\le \alpha<\beta\le M/2-1$) upon $|0^\zeta\rangle$, which generalizes  $\zeta$ states in the Hubbard model.

The two representations $\CH_\eta$ and $\CH_\zeta$ are always distinguished by a reflection operator in $\text{O}(N)$, which can be realized by $e^{i\pi n_j}$ (the site $j$ can be chosen arbitrarily). $\CH_\eta$ has parity $+1$ under $e^{i\pi n_j}$ and hence is $\text{O}(N)$ invariant. $\CH_\zeta$, on the other hand, has parity $-1$ under $e^{i\pi n_j}$ and hence is only $\SO(N)$ invariant. The fully occupied state, which is apparently a highest-weight state, has parity $(-1)^{M/2}$ under the reflection $e^{i\pi n_j}$. So it belongs to $\CH_\eta$ when $M/2$ is even and $\CH_\zeta$ when $M/2$ is odd. This also explains why the highest weight vectors of $\CH_\eta$ and $\CH_\zeta$ depend on the parity of $M/2$. When $N$ is odd,  $\CH_\eta$ and $\CH_\zeta$ can also be distinguished by the fermionic parity \eqref{eq:fermParity}. In this case, $\CH_\eta$ has fermionic parity $+1$ and $\CH_\zeta$ has fermionic parity $-1$.

Altogether, there are $\dim \CH_\eta$ linearly independent $\text{O}(N)$ singlets and $2\dim \CH_\eta$ linearly independent $\text{SO}(N)$ singlets. The explicit expression of $\dim \CH_\eta$ can be derived using Weyl dimension formula (\ref{SOdim}). Some small $M$ examples are:
\begin{align}\label{ONdim}
&M=4: 
N+1\nonumber\\
&M=6: 
\binom{N+3}{3}\\
&M=8: 
\frac{N+3}{3}\binom{N+5}{5}\nonumber
\end{align}
The general formula for the dimension of the singlet states is given by
\begin{gather}
D(N,M) = \frac{\prod\limits^{M+1-4i>0}_{i=1}\binom{N+M-1-2i}{M+1-4i}}{\prod\limits^{M+1-4i>0}_{i=1}\binom{M-1-2i}{M+1-4i}} \label{eq:gendimfor}
\end{gather}
We note that, for a fixed even $M$, the number of singlet states grows at large $N$ as $\dim \CH_\eta \sim N^{\frac{M(M-2)}{8}}$ .  If we fix $N$ and consider large $M$ we get $ \dim \CH_\eta \sim e^{0.22 N M}$.

\subsection{Energy spectrum and degeneracy \label{sec:enSpectrumAndDeg}}

For the full Hamiltonian $H=T_M+H_\mu+V_M+\widetilde H_{\text{int}}$ given by eq. \eqref{FullHam} (same when $\widetilde H_{\text{int}}$ is replaced by $H_{OT}$ \eqref{hjk}), its spectrum in $\mathbb S=\CH_\eta\oplus \CH_\zeta$ is determined by $H_\mu$
 and $V_M$. 
The spectrum of $V_M$ is particularly simple. Noticing that it is even under the reflection $e^{i \pi n_j}$,  we have $V_M \CH_\eta\subset \CH_\eta$ and $V_M \CH_\zeta\subset \CH_\zeta$. Since $\CH_\eta$ and $\CH_\zeta$ are irreducible representations of $\SO(M)$, we can use Schur's lemma to argue that the $\SO(M)$ invariant operator $V_M$ becomes a constant when restricted to either $\CH_\eta$ or $\CH_\zeta$. Using $V_M|0\rangle =0$ and $V_M|0^\zeta\rangle =-2^{1-\frac{M}{2}}N U|0^\zeta\rangle$, we conclude
\begin{align}
\left.V_M\right|_{\CH_\eta}=0, \,\,\,\,\, \left.V_M\right|_{\CH_\zeta}=-2^{1-\frac{M}{2}}N U
\end{align} 

The spectrum of the chemical potential term $H_\mu$ restricted to $\mathbb S$ is encoded in the partition function $\CZ(\beta)\equiv \tr_{\mathbb S} \, e^{-\beta \sum_\alpha \mu_\alpha h_\alpha}$, which group theoretically is equal to  the sum of two $\SO(M)$ characters 
\begin{align}
\CZ(\beta)=\chi_{\lambda_\eta}^{\SO(M)}(\bm x) + \chi_{\lambda_\zeta}^{\SO(M)}(\bm x) 
\end{align}
where $x_\alpha\equiv e^{-\beta\mu_\alpha}$. Each character can be computed using Weyl character formula (\ref{Weylchar}).  These Weyl characters admit the following expansions 
\begin{align}
\label{eq:chiOfXExpansion}
&\chi_{\lambda_\eta}^{\SO(M)}(\bm x)=\sum_{\bm p} D^\eta_{\bm p} \prod_{\alpha} x_\alpha^{p_\alpha}\nonumber\\
& \chi_{\lambda_\zeta}^{\SO(M)}(\bm x) =\sum_{\bm p}D^\zeta_{\bm p} \prod_{\alpha} x_\alpha^{p_\alpha}
\end{align}
where  $\{p_\alpha\}$ take values in all integers when $N$ is even and all half integers when $N$ is odd. The coefficients $D^\eta_{\bm p}$ ($D^\zeta_{\bm p}$) are nonnegative integers, and they vanish for all but a finite number of vectors $\bm p$. Because of the identification $x_\alpha=e^{-\beta\mu_\alpha}$, the characters implies that there are $D^\eta_{\bm p}$ ($D^\zeta_{\bm p}$) linearly independent states in $\CH_\eta$ ($\CH_\zeta$) that diagonalize all $\{h_\alpha\}$ simultaneously with eigenvalues $\{p_\alpha\}$. Altogether, the eigenenergies of the full Hamiltonian restricted to the scar subspace can be summarized as 
\small
\begin{align}\label{eq:singletsEnergies}
&\CH_\eta:\left\{\sum_{\alpha} \!-\mu_\alpha\left(p_\alpha\!+\!\frac{N}{2}\right), \,\, \text{for all}\,\, \bm p\,\, \text{satisfying}\,\, D^\eta_{\bm p}\!>\!0\right\}\nonumber\\
&\CH_\zeta:\left\{\sum_{\alpha} \!-\mu_\alpha\left(p_\alpha\!+\!\frac{N}{2}\right)\!-\!\frac{NU}{2^{\frac{M}{2}-1}}, \,\, \text{for all}\,\, \bm p\,\, \text{satisfying}\,\, D^\zeta_{\bm p}\!>\!0\right\}
\end{align}
\normalsize
For finite given $N$ and $M$ the expansion \eqref{eq:chiOfXExpansion} can be performed analytically or using analytical math software and the positive $D_{\bm p}$ can be read off from it. Therefore the spectrum in the scar subspace is known exactly analytically for the systems that we study numerically in Sec. \ref{sec:numerics} and \ref{sec:dos}.

For generic chemical potentials $\mu_\alpha$ and interaction strength $U$, the energy spacing does not have a common divisor. So we will not observe revivals starting from a generic state in $\mathbb S$.
 However, because $\{p_\alpha\}$ are integers or half-integers, revivals are possible with special choices of $\mu_\alpha$
 
  \begin{align}
  \frac{\mu_\alpha}{\mu_\beta}\in\mathbb Q, \,\,\,\,\,\, \frac{\mu_\alpha}{2NU}\in\mathbb Q
  \end{align}
where the second condition can be removed if we only consider scars in $\CH_\eta$ or $\CH_\zeta$.

When $M=4$ or $6$, $D^\eta_{\bm p}$ and $D^\zeta_{\bm p}$ are either 0 or 1, which means that $H_\mu$ does not have  any degenerate energy level   within $\CH_\eta$ or $\CH_\zeta$. On the other hand, double degeneracy happens between $\CH_\eta$ and $\CH_\zeta$ when $N$ is even. This does not happen for odd $N$ because $\CH_\eta$ and $\CH_\zeta$ are distinguished by fermionic parity when $N$ is odd. 
For instance, when $M=6$ and $N=4$, $H_\mu$ has 19 doubly degenerate energy levels. Such degeneracies are broken by $V_M$. When $M\ge 8$, $D^\eta_{\bm p}$ and $D^\zeta_{\bm p}$ can be larger than 1. It corresponds to  degeneracies within $\CH_\eta$ or $\CH_\zeta$ and hence cannot be removed by $V_M$. For example, the three states $\eta_{12}^\dagger \eta_{34}^\dagger|0\rangle$, $\eta_{13}^\dagger \eta_{24}^\dagger|0\rangle$ and $\eta_{14}^\dagger \eta_{23}^\dagger|0\rangle$ have the same quantum numbers with respect to the Cartan generators, and hence have the same energy. Indeed, when $M=8$ and $N=4$, the case studied numerically in Sec. \ref{sec:NumMlt4}, $\CH_\eta$ contains 32 triply degenerate energy levels, and one energy level with degeneracy $6$. The same degeneracies are present in $\CH_\zeta$, in agreement with numerical findings.

We conjecture that the remaining degeneracies we observe for $M>6$ are ``unbreakable," i.e. they cannot be removed by any {\it local} perturbations that preserve the decoupling of the scars.
If we allow non-local terms, than using the second term in \eqref{eq:generalHsForm} we can easily break the degeneracies. For instance, we can consider
\begin{gather}
\label{eq:Jab2}
H^{\rm nl}_2 = \sum_{A,B=1}^M r_{AB} \left(J^{AB}\right)^2\ ,
\end{gather}
where $r_{AB}$ are a set of real random numbers or
\begin{gather}
\label{eq:JabJcd}
H^{\rm nl}_m = \sum_{A,B,C,D=1}^M r_{ABCD} \left(J^{AB}J^{CD}\right)\ ,
\end{gather}
where $r_{ABCD}$ are real random numbers and the sum over $A,B,C,D$ only includes combinations where either all the four indexes are different or $A,B=C,D$ (to ensure each term is Hermitian). 

A proof of the degeneracies for local Hamiltonians may proceed along the following lines. The Hamiltonian that supports the $SO(N)$ singlet states as eigenstates must have the following form \cite{pakrouski2020GroupInvariantScars}
\begin{gather}
H = H_s + \sum_{i<j,A} O_{ij} \psi^A_i \psi^A_j\,,
\end{gather}
where $H_s$ is a singlet operator and $\sum_A\psi^A_i\psi^A_j$ is a generator of $SO(N)$. The idea is to consider all possible terms that can enter into $H_s$, and to show that they can be separated into 3 groups: a) those that preserve the degeneracies within the scar subspace and are local; b) those that preserve the degeneracies within the scar subspace and are non-local, but their non-locality can be "fixed" by adding a suitable $OT$ term; c) those that may break the scar degeneracies but are then necessarily non-local and cannot be made local by means of adding an $OT$ term. Therefore the locality of the Hamiltonian $H$ is considered as a whole while its individual summands may be non-local. The locality in our definition means in particular that there exists some number $K$ such that for any term in $H$ any two Majorana operators in it are not too far $\psi^A_i, \psi^B_j$ $|i-j| < K$. 

The most general form of $H_s$ is 
\begin{gather}
H_s = \mu_{AB} \sum_i \psi^A_i \psi^B_i + \nu_{ABCD} \sum_{i,j} \psi^A_i \psi^B_i  \psi^C_j \psi^D_j + \ldots+\notag\\
\lambda \sum_{i,\ldots,j}\epsilon_{A_1 A_2\ldots A_{M-1} A_M}  \psi^{A_1}_i \psi^{A_2}_i \ldots  \psi^{A_{M-1}}_j \psi^{A_M}_j\,,\label{eq:generalHsForm}
\end{gather}
where we group terms by the (even) number of contributing Majorana operators and where $\nu_{ABCD} $ is an arbitrary tensor and 
$\epsilon_{A_1 A_2\ldots A_{M-1} A_M}  \psi^{A_1}_i \psi^{A_2}_i \ldots  \psi^{A_{M-1}}_j \psi^{A_M}_j$ is the fully antisymmetric tensor. We can consider these terms separately.

The first term, the chemical potential is obviously local, since only the fermions on the same site interact. 
The second and last term in \eqref{eq:generalHsForm} are obviously non-local. For the last term in \eqref{eq:generalHsForm} however we can add such a combination of $OT$ terms that reduces it to the local $V_M$ defined in \eqref{eq:allFlavourUakaVsubM}. While for the second term through exhaustive search (considering the two-site, four-Majorana operators with $M=6$), we found that there is no $OT$ term capable of removing its non-locality for arbitrary $\nu_{ABCD}$. At present, we do not have the complete proof for other terms, which contain $6$ or more Majorana operators, but we conjecture that there exists no such $OT$ term that could make these intermediate terms in $H_s$ local.

In summary, all the terms in $H_s$ but first and last are eliminated by the locality requirement and we see that any local $H_s$ is a sum of an element of Lie algebra $\mathfrak{so}(M)$ (chemical potential term) and generalized Hubbard interaction $V_M$ (only shifts scar energy by a constant). This leads us to conjecture that the scar subspace spectrum is fully determined by this first chemical potential term as long as the Hamiltonian is required to be local.  Its spectrum can be computed using the Weyl character formula \cite{humphreys2012introduction} and therefore contains certain (``unbreakable'') degeneracies.

\subsection{Product scar states}
Product states are very special because their entanglement entropy vanishes.  In  $\mathbb{S}$, product states are either empty or fully filled for each of the $M/2$ flavors.  Hence there are $2^{M/2}$ such states. To describe their wavefunctions, we define the following $N$-fermion operators 
\begin{align}
\mathcal A^\dagger_\alpha\equiv c^{\dagger}_{1\alpha} c^{\dagger}_{2\alpha}\cdots c^{\dagger}_{N\alpha}, \,\,\,\,\, \alpha=1,2,\cdots,M/2
\end{align}
Then the $2^{M/2}$ product states can be expressed as 
\begin{align}
\label{eq:c1productSinglets}
|\alpha_1,\cdots, \alpha_\kappa\rangle = \CA^\dagger_{\alpha_1}\cdots \CA^\dagger_{\alpha_\kappa} |0\rangle
\end{align}
where  $1\le \alpha_1<\alpha_2<\cdots< \alpha_\kappa\le M/2$. In particular, $\kappa=0$ corresponds to the empty vacuum $|0\rangle$, and $\kappa=M/2$ corresponds to the fully occupied state. Since $\CA^\dagger_\alpha$ has odd parity under the reflection $e^{i\pi n_j}$, the product state $|\alpha_1,\cdots, \alpha_\kappa\rangle$ belongs to $\CH_\eta$ when $\kappa$ is even and $\CH_\zeta$ when $\kappa$ is odd. For the full Hamiltonian $H$, c.f. eq. (\ref{FullHam}), $|\alpha_1,\cdots, \alpha_\kappa\rangle$ is an eigenstate with energy
\begin{align}
E_{\alpha_1\cdots \alpha_\kappa}=-N\sum_{m=1}^\kappa\mu_{\alpha_m}+\frac{(-1)^\kappa-1}{2^{\frac{M}{2}}} NU
\end{align}

In the case of $M=4$, the product states in $\mathbb S$ are 
\begin{align}\label{prodst}
&\kappa=0: \,\, |0\rangle,\,\,\,\,\,\,\, \kappa=2:\,\, |1,2\rangle =\CA_1^\dagger \CA_2^\dagger|0\rangle\nonumber\\
&\kappa=1: \,\, |1\rangle=\CA_1^\dagger |0\rangle,\,\,\,\,\,\,\, |2\rangle = \CA_2^\dagger|0\rangle
\end{align}
where the $\kappa=0$ and $\kappa=2$ states are $\eta$ states  $|0^\eta\rangle$  and $|N^\eta\rangle$ (c.f. eq. (\ref{etastate})), and the two $\kappa=1$ states are $\zeta$ states $|0^\zeta\rangle$  and $|N^\zeta\rangle$ (c.f. eq. (\ref{zetastate})), which have  total spin $\pm \frac{N}{2}$.

\subsection{The $M=6$ case}\label{m6case}

When $M=6$, the $\text{O}(N)$ invariant subspace $\CH_\eta$ has dimension $\binom{N+3}{3}$. Consider states in $\CH_\eta$ of the form $(\eta_{12}^\dagger)^{k_{12}} (\eta_{13}^\dagger)^{k_{13}} (\eta_{23}^\dagger)^{k_{23}} |0\rangle $. First, by a direct computation, we find that $(\eta_{12}^\dagger)^{k_{12}} (\eta_{13}^\dagger)^{k_{13}} (\eta_{23}^\dagger)^{k_{23}} |0\rangle $ has the norm $\frac{N! k_{12}!k_{13}!k_{23}!}{(N-k_{12}-k_{13}-k_{23})!}$ and hence is nonvanishing when $k_T\equiv k_{12}+k_{13}+k_{23}\le N$. Next, these states are linearly independent, because $\{k_{12},k_{13},k_{23}\}$ uniquely fixes  the fermion numbers of the three flavors, namely $k_{12}+k_{13}$ particles of flavor 1, $k_{12}+k_{23}$ particles of flavor 2 and $k_{13}+k_{23}$ particles of flavor 3. Finally, counting nonnegative integer solutions of the inequality  $k_{12}+k_{13}+k_{23}\le N$ precisely reproduces $\binom{N+3}{3}$. Therefore, an orthonormal basis of $\CH_\eta$ that diagonalizes the three Cartan generators $\{h_1, h_2, h_3\}$ simultaneously is 
\begin{align}\label{normalscar}
&|k_{12},k_{13},k_{23}\rangle =  C_{\bm k} (N)\prod_{1\le\alpha<\beta\le 3}(\eta^\dagger_{\alpha\beta})^{k_{\alpha\beta}} |0\rangle, \,\,\,\,\, k_T\le N
\end{align}
where the normalization factor $C_{\bm k}(N)$ is given by
\begin{align}
C_{\bm k}(N)=\sqrt{\frac{(N-k_T)!}{N! k_{12}!k_{13}!k_{23}!}}
\end{align}
These states \eqref{normalscar} are also constructed in \cite{Nakagawa:2022jsg} as energy eigenvectors of the $\SU(3)$ Hubbard model.
The eigenenergy of $|k_{12},k_{13},k_{23}\rangle$ with respect to our full Hamiltonian $H$ is 
\begin{align}
E_\eta(k_{12},k_{13}, k_{23})=-\sum_{i<j} k_{ij}(\mu_i+\mu_j)
\end{align}

For the $\SO(N)$ invariant subspace $\CH_\zeta$, we can construct an orthonormal basis similarly
\begin{align}\label{psibasis}
&|k_{12}, k_{13},k_{23}\rangle^\zeta= \notag\\ &C_{\bm{k}}(N)(\eta_{12}^\dagger)^{k_{12}} (\zeta_{13}^\dagger)^{k_{13}} (\zeta_{23}^\dagger)^{k_{23}} |0^\zeta\rangle, \,\,\, k_T \leq  N 
\end{align}
They are eigenstates of the  Hamiltonian (\ref{FullHam}) with energy 
\begin{align}
E_\zeta(k_{12},k_{13}, k_{23})=-\sum_{i<j} k_{ij}(\widetilde\mu_i+\widetilde\mu_j) +N\widetilde\mu_3-\frac{NU}{2^{\frac{M}{2}-1}}
\end{align}
where $\widetilde{\mu}_{1,2} = \mu_{1,2}, \widetilde{\mu}_3 = -\mu_3$.

With the explicit wavefunctions of all the $\SO(N)$ singlets in the $M=6$ case, we can calculate their  entanglement entropy analytically (more details can be found in the appendix \ref{sec:m6entropy}).
We will focus on $\CH_\eta$ but the same method works for $\CH_\zeta$. A similar calculation was performed for the $\eta$ states in usual Hubbard model \cite{Vafek_2017}, and for a special class of $\eta$ states in multi-component Hubbard model in \cite{Nakagawa:2022jsg}.  In the thermodynamic limit we find that the entropy $S_{N_1}(\bm k)$ scales as the {\it logarithm} of $N_1$, the number of sites in the sub-system $\Sigma_1$
\begin{align}
S_{N_1}(k_{12},k_{13},k_{23})
\sim\frac{3}{2}\log (N_1)\label{Sscale}\,,
\end{align}
where $k=k_{12}+k_{23}+k_{13}$. This result does not depend on the dimensionality of the original system. The coefficient $\frac{3}{2}$ for the logarithmic behavior agrees with a calculation in \cite{Nakagawa:2022jsg}. Let us note that the coefficient 
$\frac{3}{2}$
applies only to the most typical $\eta$-states where $k_{12}$, $k_{13}$ and $k_{23}$ are all large. If only two of them are large, then the coefficient of $\log (N_1)$ is reduced to $1$; if only one of them is large, then it is reduced to $1/2$.

\subsection{The structure of singlets for $M\ge 8$}
When $M=8$, there are six different $\eta_{\alpha\beta}^\dagger$. In order to construct explicit wavefunctions of $\CH_\eta$, we consider the following set of states, which generalizes the $M=6$ case, 
\begin{align}\label{type1}
V_\eta^{\rm \rom 1}\equiv \text{Span}\left\{\prod_{\alpha<\beta}\left(\eta^\dagger_{\alpha\beta}\right)^{k_{\alpha\beta}}|0\rangle, \,\,\,\sum_{\alpha<\beta} k_{\alpha\beta}\le N\right\}.
\end{align}
A simple counting yields  $\dim V_\eta^{\rm \rom 1}=\binom{N+6}{6}$, which is smaller than $\frac{N+3}{3}\binom{N+5}{5}$, i.e. the dimension of $\CH_\eta$ when $M=8$. It means that $V_\eta^{\rm\rom 1}$  is only a subset of $\CH_\eta$. For example, the fully filled state $|\bar 0\rangle=\prod_{\alpha}^4\prod_{j=1}^N c_{j\alpha}^{\dagger}|0\rangle$ does not belong to $V_\eta^{\rm\rom 1}$. Noticing that $|\bar 0\rangle$ is actually the highest-weight state of $\CH_\eta$, we build upon it another set of states 
\begin{align}\label{type2}
V_\eta^{\rm\rom 2}\equiv \text{Span}\left\{\prod_{\alpha<\beta}\left(\eta_{\alpha\beta}\right)^{\ell_{\alpha\beta}}|\bar 0\rangle, \,\,\, \sum_{\alpha<\beta} \ell_{\alpha\beta}<N\right\}
\end{align}
which has dimension $\binom{N+5}{6}$. Since states in $V_\eta^{\rm \rom 1}$ have total fermion number $Q\le 2N$ and states in $V_\eta^{\rm \rom 2}$ have total fermion number $Q> 2N$, the two sets $V_\eta^{\rm \rom 1}$ and $V_\eta^{\rm \rom 2}$ have no intersection. Adding up their dimensions gives exactly the dimension of $\CH_\eta$. Therefore  an orthonormal   basis of $\CH_\eta$ is 
\begin{align}
\prod_{\alpha<\beta}\left(\eta^\dagger_{\alpha\beta}\right)^{k_{\alpha\beta}}|0\rangle, \,\,\,\,\, \prod_{\alpha<\beta}\left(\eta_{\alpha\beta}\right)^{\ell_{\alpha\beta}}|\bar 0\rangle
\end{align}
where  $k_{\alpha\beta}$ and $\ell_{\alpha\beta}$ satisfy $\sum_{\alpha<\beta}k_{\alpha\beta}\le N, \,\,\,\,\, \sum_{\alpha<\beta}\ell_{\alpha\beta}< N$.

Similarly, one can show that $\CH_\zeta$ is an orthonormal  direct sum  of $V_\zeta^{\rm \rom 1}$ which is spanned by
\begin{align}
\prod_{1\le \alpha<\beta\le 3}\left(\eta^\dagger_{\alpha\beta}\right)^{k_{\alpha\beta}}\prod_{1\le \gamma \le 3} \left(\zeta^\dagger_{\gamma4}\right)^{k_{\gamma_4}}|0^\zeta\rangle, \,\,\,\sum_{\alpha<\beta} k_{\alpha\beta}\le N
\end{align}
and $V_\zeta^{\rm \rom 2}$ which is  spanned by 
\begin{align} 
\prod_{1\le \alpha<\beta\le 3}\left(\eta_{\alpha\beta}\right)^{\ell_{\alpha\beta}}\prod_{1\le \gamma \le 3} \left(\zeta_{\gamma4}\right)^{\ell_{\gamma_4}}|\bar 0^\zeta\rangle, \,\,\,\sum_{\alpha<\beta} \ell_{\alpha\beta}< N
\end{align}
where $|\bar 0^\zeta\rangle=\prod_{\alpha=1}^3\prod_{j=1}^N c^\dagger_{j\alpha} |0\rangle$ is the highest-weight state of $\CH_\zeta$. The basis of $V_\eta^{\rm \rom 1},V_\eta^{\rm \rom 2},V_\zeta^{\rm \rom 1}$ and $V_\zeta^{\rm \rom 2}$ are eigenstates of the Hamiltonian $H$, and their corresponding eigenenergies can be easily derived by reading off the fermion number of each flavor. For example, the energy of $\prod_{\alpha<\beta}\left(\eta^\dagger_{\alpha\beta}\right)^{k_{\alpha\beta}}|0\rangle$ is 
\begin{align}
E_\eta(\bm k)=-\sum_{\alpha<\beta}k_{\alpha\beta}(\mu_\alpha+\mu_\beta)
\end{align}
We will compute their entanglement entropy numerically in the next section.

For $M\ge 10$, it becomes very hard to write down explicit wavefunctions of all $\SO(N)$ singlets.

In \cite{Nakagawa:2022jsg}, some special states belonging to $\CH_\eta$ are considered, namely $(\eta_{12}^\dagger)^{k_{2}}\cdots (\eta_{1,M/2}^\dagger)^{k_{M/2}}|0\rangle$ with $k_2+\cdots+ k_{M/2}\le N$. They are eigenstates of the generalized Hubbard potential (\ref{SUNHubbard}) and their entanglement entropy can be evaluated analytically. In the thermodynamic limit, the entanglement entropy is found to scale as $S_{\Sigma_1}\sim\frac{M-2}{4}\log N_1$ \cite{Nakagawa:2022jsg}. When $M=6$, this expression gives $S_{\Sigma_1}\sim\log N_1$,  different from what we found in eq. (\ref{Sscale}). The difference is because $S_{\Sigma_1}\sim\log N_1$ corresponds to the entanglement entropy of $\eta$ states $(\eta^\dagger_{12})^{k_{12}}(\eta_{13}^\dagger)^{k_{13}}|0\rangle$, which do not involve $\eta^\dagger_{23}$. These special $\eta$ states are not captured by the analysis in section \ref{m6case}, since the thermodynamic limit there requires all three $\frac{k_{ij}}{N}$ to be finite as $N\to\infty$.

\subsection{Off-Diagonal Long-Range Order}

By construction, the singlet states have long-range correlations, and we can confirm this by studying the spatial dependence of the following operators
\begin{gather}
\label{eq:ODLROdef}
O^{AB}_{ij}(s) = \langle s| \psi^A_i \psi^B_i \psi^B_j \psi^A_j| s \rangle, 
\end{gather}
It is easy to check that if $i \neq j$ this operator is real and does not depend on the indices $i$ and $j$. Indeed, let us consider a rotation $Q_{ik}$  
that acts in the $(i,k)$ plane and leaves the rest of the vectors untouched. Then it is easy to check that
\small
\begin{gather}
O^{AB}_{ij}(s)=\langle s| \psi^A_i \psi^B_i \psi^B_j \psi^A_j| s \rangle =\langle s| Q^{-1}_{ik} \psi^A_i \psi^B_i \psi^B_j \psi^A_j Q_{ik }| s \rangle = \notag\\
=\langle s| \psi^A_k \psi^B_k \psi^B_j \psi^A_j| s \rangle = O^{AB}_{kj}(s) = O^{AB}(s),
\end{gather}
\normalsize
the value of this operator does not depend on the spatial indexes. Therefore the associated correlations do not decay and remain finite and constant at arbitrary distances! The expectation value \eqref{eq:ODLROdef} depends on the choice of flavors $AB$ and the singlet state $\ket{s} \in \mathbb{S}$. But if we sum over $A$ and $B$ we would get a simpler operator
\begin{gather}
O(s) = \sum_{A \neq B} O^{AB}(s) = \sum_{A\neq B} 
\langle s| \psi^A_i \psi^B_i \psi^B_j \psi^A_j| s \rangle \notag\\
=-
\sum_{A\neq B}  \langle s| \psi^A_i \psi^A_j  \psi^B_j \psi^B_i| s \rangle 
\end{gather}
where we have used anticommutation relations. After that we can get
\begin{gather}
O(s)=
-\sum_{A, B}  \langle s| \psi^A_i \psi^A_j  \psi^B_j \psi^B_i| s \rangle +\sum_{A=B}  \langle s| \psi^A_i \psi^A_j  \psi^B_j \psi^B_i| s \rangle \notag\\
= \sum_{A=B}  \langle s| \psi^A_i \psi^A_j  \psi^B_j \psi^B_i| s \rangle = \frac{M}{4} \label{eq:sumodlro},
\end{gather}
where we have used the fact that $|s\rangle$ is annihilated by the action of hopping $T_{ij} = \sum_A \psi^A_i \psi^A_j$ and the last equality follows from the anti-commutation relations \eqref{eq:antiCommEq}.

Another ``sum rule" arises if we 
average $O^{AB}_{ij}$ over all the singlet states
\begin{gather}
\label{eq:ODLRO2ndSumRule}
\widetilde{O}^{AB} = \frac{1}{\dim \mathbb S}\sum_{\ket{s} \in \mathbb{S}} O^{AB}_{ij}(s)
\end{gather}
which amounts to computing the trace of $ \psi^A_i \psi^B_i \psi^B_j \psi^A_j  $ over the scar subspace. When $i\not=j$, its value is independent of the choice of $i, j$. For $A\not=B$, we have 
\begin{align}\label{trS}
\sum_{i, j}\frac{\tr_{\mathbb S}( \psi^A_i \psi^B_i \psi^B_j \psi^A_j  )}{\dim \mathbb S}=N(N-1)\widetilde{O}^{AB}+\frac{N}{4}
\end{align}
where $\dim\mathbb S=2\dim\mathcal H_\eta$ is the dimension of scar sector (see eq. (\ref{ONdim}) for explicit expressions of $\dim\mathcal H_\eta$ when $M$ is small). On the L.H.S of eq. (\ref{trS}), the double sum over $i, j$ yields $J^{AB}J^{BA}$. Noticing that  $\tr_{\mathbb S} (J^{AB}J^{BA})$ is actually independent of $A, B$, we can replace it by $\sum_{A<B}\tr_{\mathbb S} (J^{AB}J^{BA})/(\frac{1}{2}M(M-1))$. The sum over $A,B$ leads to the Casimir of $\SO(M)$ and the latter is a constant in the scar subspace with its value  given by $\frac{MN}{8}(M+N-2)$. Altogether, we obtain the value of  $\widetilde{O}^{AB}$:
\begin{align}
\widetilde{O}^{AB}=\frac{1}{4(M-1)}
\end{align}
which implies that  $O^{AB}(s)$ is nonvanishing for at least one $|s\rangle$.

\section{A bound on the entanglement entropy of singlet states \label{sec:entrBound}}

An upper bound on the entanglement entropy of singlet states can be obtained without knowing the explicit wavefunctions. We will derive the bound for $\eta$ states and it will be easy to see that the same bound also holds for $\zeta$ states. As noted earlier, we may divide the full lattice into two disjoint sublattices $\Sigma_1\cup\Sigma_2$, with $N_a$ sites in  $\Sigma_a$. Let $N_1$ be much smaller than $ N_2=N-N_1$, but still a large number in the thermodynamic limit. On each sublattice $\Sigma_a$, we can construct its own $\eta$ operators $\eta^{a\dagger}_{\alpha\beta}$  that create O$(N_a)$ singlets upon the empty vacuum $|0\rangle_a$. We use $\CH^a_\eta$ to denote the sub-Hilbert space spanned by all O$(N_a)$ singlets  on $\Sigma_a$.  The dimension of $\CH^a_\eta$ is given by $D(N_a, M)$, c.f. eq. (\ref{eq:gendimfor}). Recall that an O$(N)$ invariant state  is constructed by acting with $\eta^\dagger_{\alpha\beta}$ on $|0\rangle$. Because $\eta^\dagger_{\alpha\beta}=\sum_{a=1}^2\eta^{a\dagger}_{\alpha\beta}$ and $|0\rangle= |0\rangle_1\otimes |0\rangle_2$, the  O$(N)$ invariant state should also belong to $\CH^1_\eta\otimes\CH^2_\eta$. Applying the Schmidt decomposition to this tensor product yields that for any $|s\rangle \in\CH_\eta$, we have 
\begin{gather}
\ket{s} = \sum_{n=1}^{D(N_1, M)} \Lambda_n \ket{s^1_n} \otimes \ket{s^2_n},\,\,\,\,\,\sum_{n=1}^{D(N_1, M)} \Lambda_n^2 = 1\ , \label{eq:singletdecom}
\end{gather}
where $\ket{s_n^a}$ are orthonormal states in $\mathcal{H}_{\eta}^a$ and are most importantly O$(N_a)$ invariant. We observe that the number of the non-zero terms in the decomposition \eqref{eq:singletdecom} is significantly reduced compared to the full dimension of the sub-system $\Sigma_1$ ($2^{N_1M/2}$). Then it is clear that the entanglement entropy of $|s\rangle$ in the subsystem $\Sigma_1$ is bounded by $\log D(N_1, M)$ from above
\begin{gather}
\label{entropybound}
S_{\rm EE} \leq \log D(N_1,M).
\end{gather}  
For the $\text{O}(N)$ singlets we can use the formula \eqref{eq:gendimfor} and its asymptotic expansion in the thermodynamic limit $D(N_1, M)\sim N_1^{\frac{M(M-2)}{8}}$
which leads to the bound
\begin{gather}
S_{\rm EE} \lesssim \frac{M(M-2)}{8} \log N_1~.
\end{gather}

Based on our explicit calculations, such as \eqref{Sscale}, we conjecture that the maximum entropy of the scar states in the thermodynamic limit instead grows as 
\begin{gather}
\label{eq:entrMaxConj}
S_{\rm EE} \rightarrow \frac{M(M-2)}{16} \log N_1\ ,
\end{gather}
which suggests that the bound (\ref{entropybound}) is far from being saturated.

We can generalize the argument above to other systems with an action $\rho$ of some  group $G$ on the total Hilbert space $\CH$. We assume that $G$ has two subgroups, $G_1$ and $G_2$, satisfying the following conditions:
(i) $G_1 \times G_2 \subset G$, 
(ii) $G_1 \cap G_2 = \{e\}$, and
(iii) the restriction of $\rho$ to each $G_a$ gives a representation of $G_a$ on the Hilbert space $\CH^a$ of the subsystem supported on  $\Sigma_a$. In each $\CH^a$, there is a subspace $\CH_0^a$ consisting of $G_a$ singlets. We denote the dimension of $\CH_0^a$ by $D(N_a)$ and assume that $D(N_1)\leq D(N_2)$. The projection of $\CH^a$ to $\CH_0^a$ can be implemented by the projector $P_{a} \equiv \int_{G_a} dg_{a}\, \rho(g_a)$, where $dg_a$ is the normalized Haar measure on $G_a$.
Now let's consider  a generic $G$-singlet $|s\rangle\in\CH$. Because of the tensor product structure $\CH=\CH^1\otimes\CH^2$, $|s\rangle$ can be expressed schematically as $|s\rangle = \sum_\kappa |\psi^1_\kappa\rangle \otimes |\psi^2_\kappa\rangle $, where $|\psi^a_\kappa\rangle\in\CH^a$.  Applying the projectors $P_1$ and $P_2$ to $|s\rangle$ yields $P_1 P_2|s\rangle = \sum_\kappa P_1|\psi^1_\kappa\rangle \otimes P_2|\psi^2_\kappa\rangle $.
The LHS is just equal to $\ket{s}$, while on the RHS $P_{1,2}$ projects onto the singlet subspace $\CH_0^{1,2}$. Therefore $|s\rangle$ belongs to $\CH_0^1\otimes \CH_0^2$. Using the Schmidt decomposition for this tensor product, we get a generalization of eq. (\ref{eq:singletdecom})
\begin{gather}
\ket{s} = \sum\limits_{n=1}^{D(N_1)} \Lambda_n \ket{s^1_n} \otimes \ket{s^2_n},\,\,\,\,\,\sum\limits_{n=1}^{D(N_1)} \Lambda_n^2 = 1\ , \label{eq:singletdecom2}
\end{gather}
where $\ket{s_n^a}$ are orthonormal states and singlets in $\mathcal{H}^a$. In other words, cutting a singlet state we always get two subsystem states that are also singlets and this condition significantly reduces the number of non-zero singular values contributing to the sum. 
This means that the entanglement entropy of any $G$-invariant state over $\Sigma_1$ is bounded by 
\begin{gather}
\label{eq:GinvBound}
S^G_{\text{EE}}  \leq \log(D(N_1)) ~.
\end{gather}

The bound \eqref{eq:GinvBound} actually allows us to conclude that the $G$-invariant many-body scars that appear in models built according to the general $H_0+OT$ prescription \cite{pakrouski2020GroupInvariantScars} always have the entanglement entropy that is significantly lower than that of generic excited states. 
From analogous bound for the typical excited state we can conclude that  $S_{\rm EE} \leq \log 
\dim \mathcal{H}^1\sim N_1$. On the other hand, the bound on the entropy of the group invariant many-body scars is \eqref{eq:GinvBound}.
If we assume that these bounds are parametrically correct and that $D(N_1) \ll \dim \mathcal{H}^1$,  then the entanglement entropy of singlet states  is parametrically smaller than that of the generic states.

Most of the many-body scars reported in the literature are characterized by anomalously low entanglement entropy. We conjecture that this can be explained by the same mechanism. The scars have a decomposition similar to \eqref{eq:singletdecom2} and the dimension of the subsystem states compatible with the scar states structure is significantly smaller than the full subsystem dimension which leads (see eq. \eqref{eq:GinvBound}) to lower entanglement. A special case of this situation is when cutting a scar necessarily gives scar states in the subsystems which is true for group-invariant scars \cite{pakrouski2020GroupInvariantScars}.

Finally we note that our entanglement scaling for the $\eta$ states in the single-band case is in agreement with the results obtained by the method \cite{entanglementBoundPRR2022} where local correlation matrix is analysed numerically to deduce a bound on the entanglement entropy of a quantum state. It has also been noticed \cite{Moudgalya:2022nll} that the same method \cite{entanglementBoundPRR2022} is useful for bounding the entanglement of the many-body scars built using the Shiraishi-Mori approach \cite{Shiraishi2017ScarsConstruction}.
% \cite{entanglementBoundPRR2022}: given a state one first chooses the allowed operator (Li) range (say, 3-site). Then one makes a "correlation" matrix where <LiLj> - <Li><Lj> is stored. Now it is shown that if the degeneracy of the 0-eigenvalue subspace is Da then the entropy of the state is bound by log(Da). They provide an algorithm for finding Da numerically. In the supplementary they show results for the eta states but do not provide any details how they are obtained.

\section{Numerical results \label{sec:numerics}}

To test our analytical predictions numerically we implement the Hamiltonian \eqref{FullHam} that was argued in Sec. \ref{erere} to support singlet many-body scars

\begin{align}\label{eq:fullHNumerics}
H=&-\sum_{\alpha=1}^{M/2} \mu_\alpha \sum_j n_{j\alpha} + U \sum_{j}i^{M/2}\psi^1_j\psi^2_j\cdots\psi_j^M \nonumber \\
&+it \sum_{\langle j,k\rangle}T_{jk}+ p H_{SB}-\frac{N}{2^{\frac{M}{2}}} U\ ,
\end{align}
where the auxiliary symmetry-breaking term $H_{SB}$ is given by either a 3-body ($H_{SB} = \widetilde H_{\rm int}$ from eq. \eqref{eq:TOT}) or 2-body ($H_{SB} = H_{\rm OT}$ from eq. \eqref{hjk}) interactions. The hopping strength is set to $t = 1$.

The values of chemical potentials $\mu_\alpha$ and parameters $p$ and $U$ will be specified for each particular system size and were chosen to ensure the best visual presentation of the entanglement entropy plots. While our construction guarantees the presence of many-body scars in any dimension and is insensitive to the boundary conditions we perform the numerical calculations in 1D and use open boundary conditions.

Given a Hamiltonian we perform full exact diagonalization obtaining the values for all the eigenvalues and eigenvectors. 
As the first test we examine the dimension of the $\SO(N)$-singlet subspace. To do this we scale the part ($T_M + H_{SB}$) of the Hamiltonian \eqref{eq:fullHNumerics} that is proportional to $\SO(N)$ generators. Then we count the number of the energy levels that remain unchanged up to numerical precision. The number of these states is given in the Table \ref{tab:singletsCounting} for several system sizes and agrees with the number of $\SO(N)$ singlets given in eq. \eqref{ONdim}.

  \begin{table}[t!]
  	\centering
  	\caption{ The number of states in spectrum identified as many-body scars and the number of degenerate scar states. \label{tab:singletsCounting}
  }	
  	\label{tab:Counts}
  		\begin{tabular}{|c|c|c|c|}
  			\hline
  						 $N$	& $M$		& scars			& degenerate scars \\
			\hline
	 					2 	& 12			& 924			&	380					\\
			\hline
	 					3 	& 12			& 8448			&	6144					\\
			\hline
	 					4 	& 6			& 70				&	0					\\
			\hline
	 					4 	& 8			& 588			&	204					\\
			\hline
	 					6 	& 6			& 168			&	0				\\
			\hline
	 					8 	& 4			& 18				& 0					\\
			\hline
	 					9 	& 4			& 20				&		0			\\
	 		\hline

  		\end{tabular}
	
  \end{table}

The many-body scars are expected to stand out by violating the eigenstate thermalisation hypothesis in that an observable measured in these states clearly deviates from the thermal average at the same energy or temperature. We use entanglement entropy as one of the observables of interest.

In addition to the entanglement entropy we will also study the statistics of the level spacings in the spectrum. Going through the full spectrum we determine the level spacing and level spacing ratio
\begin{align}
\label{eq:siAndRi}
&s_i = E_i - E_{i-1}, \quad r_i = \frac{\min(s_i,s_{i+1})}{\max(s_i,s_{i+1})}
\end{align}
for every pair of energy levels.

Mean values for this level spacing ratio is known analytically \cite{Atas_2013} for certain types of random matrices: $\left< r \right> \approx 0.5359$ for the generalized orthogonal ensemble (GOE, real matrices) and $\left< r \right> \approx 0.6027$ for the generalized unitary ensemble (GUE, complex matrices).

\subsection{$M=4$ \label{sec:NumM4}}
The $M=4$ case discussed in Sec. \ref{sec:deformedHubbardM4} has the same Hilbert space as the spin-1/2 electron models studied in Refs. \cite{pakrouski2020GroupInvariantScars,Pakrouski:2021jon}. 

We study numerically the system with $N=8$, $M=4$ and set $\mu_1=0.94854$, $\mu_2=0.14631$, $U=0.72431$. The strength of the 3-body symmetry-breaking term \eqref{eq:TOT} is $p=1.25196$. Note that the model we are studying with $M=4$ reduces to the standard Hubbard model upon identification $\mu_\uparrow = \mu_1+\frac{U}{2}$; $\mu_\downarrow = \mu_2+\frac{U}{2}$ and the energies of all the scars are specified at the end of Sec. \ref{sec:deformedHubbardM4}.

\begin{figure}[htp!]
	\begin{center}
			\includegraphics[width=\columnwidth]{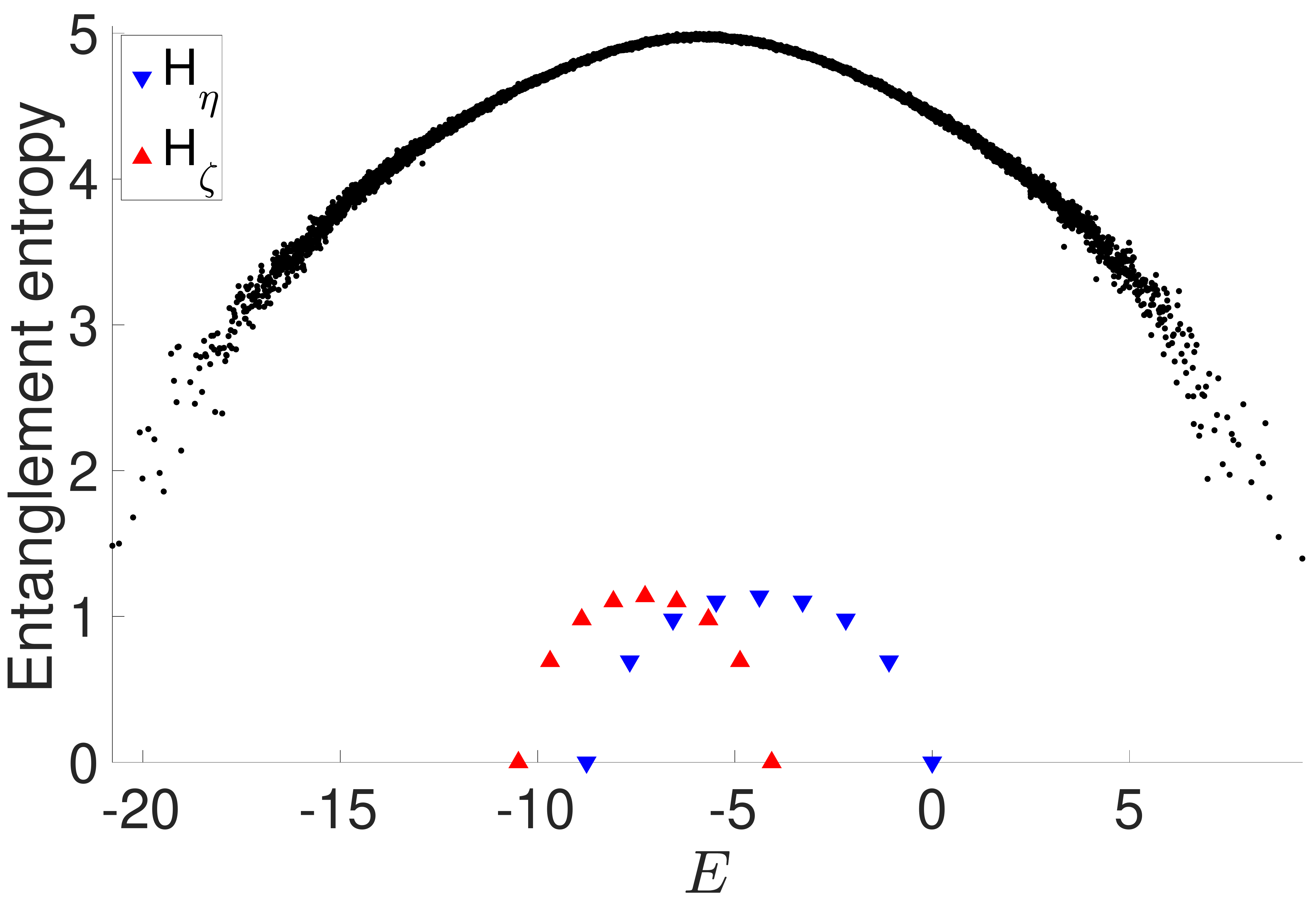}\\
			\includegraphics[width=\columnwidth]{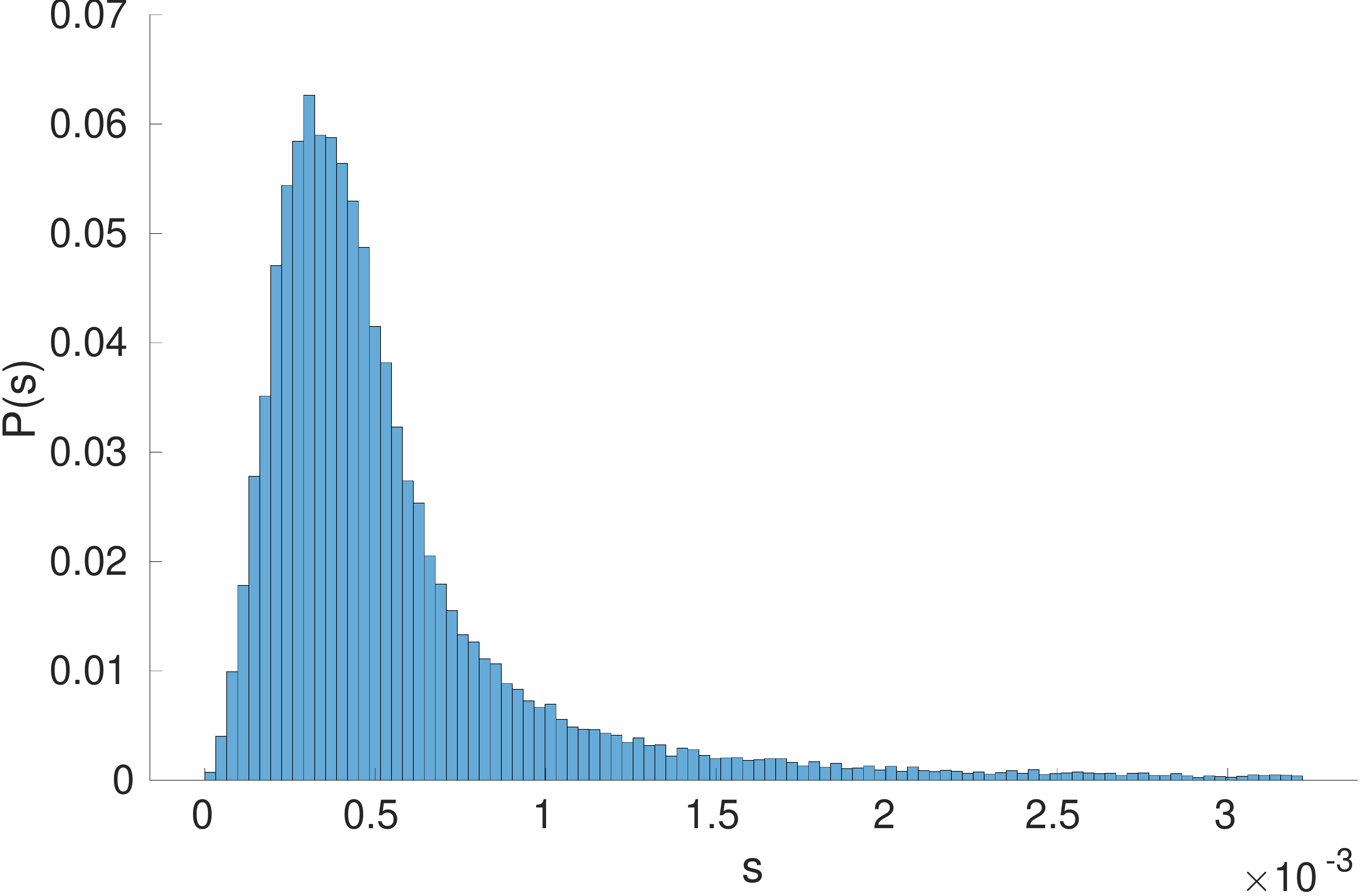}
	\end{center}
\caption{\label{fig:M4entrAndHist} Numerical results for $N$ = 8, $M$ = 4. Top panel: Entanglement entropy in every eigenstate in the sector with even fermion number, with the cut made in the middle of the 1D lattice. The $\eta$-states are shown in blue, and the $\zeta$-states in red. Bottom panel: Probability for the level spacings in the even sector. We excluded 3 percent of largest gaps from the plot and from the total norm.}
\end{figure}

The entanglement entropy in the full Hilbert space and the level spacing histogram in the even sector of the fermion number parity \eqref{eq:fermParity} are shown in Fig. \ref{fig:M4entrAndHist}. Both families of scars have the entanglement entropy significantly lower than the generic states at the same temperature. Two states ($m=0$ and $m=N$) in each scar family are product states and have exactly zero entanglement entropy.

Level spacings are analyzed in the even sector of the fermionic number parity, they are qualitatively the same in the odd sector. The average ratio of $\left< r \right> \approx 0.59949$ we obtain is very close to the GUE value. The histogram of level spacings (excluding the singlet energy levels) shown in the bottom panel of Fig. \ref{fig:M4entrAndHist} shows that the near-zero gaps are almost absent in the spectrum as is expected due to level repulsion and absence of residual symmetries. Altogether, we observe a chaotic bulk spectrum and the singlet states clearly violating ETH; this confirms that the singlet states are the many-body scars in this system.

\subsection{$M>4$ \label{sec:NumMlt4}}

The first example we consider with $M>4$ is $N=4$, $M=6$ for which a number of results are obtained analytically in Sec. \ref{m6case}. We use the chemical potentials $\mu_1= 5.69123$, $\mu_2=0.87786$, $\mu_3 = 2.50648$ and $U=2.89722$. The strength of the 3-body term \eqref{eq:TOT} is $p=0.62598$.

\begin{figure}[htp!]
	\begin{center}
			\includegraphics[width=\columnwidth]{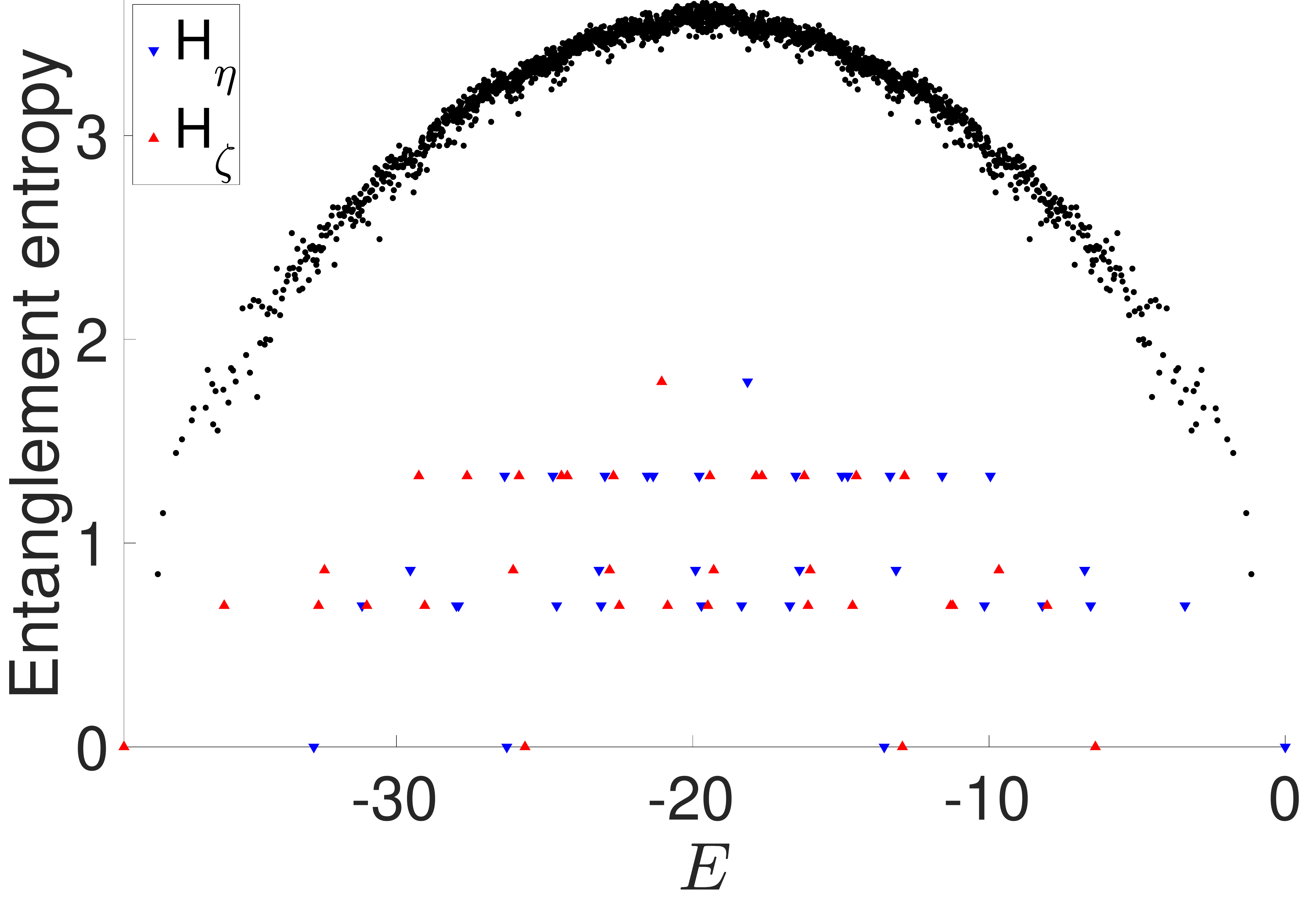}
	\end{center}
\caption{ Entanglement entropy plot for $N$ = 4, $M$ = 6 in the even fermion parity sector. The $\eta$-states are shown in blue, and the $\zeta$-states in red.
\label{fig:4x3} }
\end{figure}

The entanglement entropy in every eigenstate of the system is shown in Fig. \ref{fig:4x3}. Together with the chaos properties of the bulk spectrum (not shown) it confirms the presence of many-body scars. An interesting property rarely observed in literature \cite{multitowerScars2022,PhysRevResearch.2.043305} and also seen in the next example is that the scars are not equidistant in energy while the Hamiltonian \eqref{eq:fullHNumerics} we are studying is purely local. 

Our main multi-flavor example is $N=4$, $M=8$ where we use $\mu_1= 5.69123$, $\mu_2=0.87786$, $\mu_3 =2.50648$, $\mu_4=4.92193$ and $U=2.89722$. The strength of the 2-body symmetry-breaking term \eqref{hjk} is $p=0.62598$.

\begin{figure}[htp!]
	\begin{center}
			\includegraphics[width=\columnwidth]{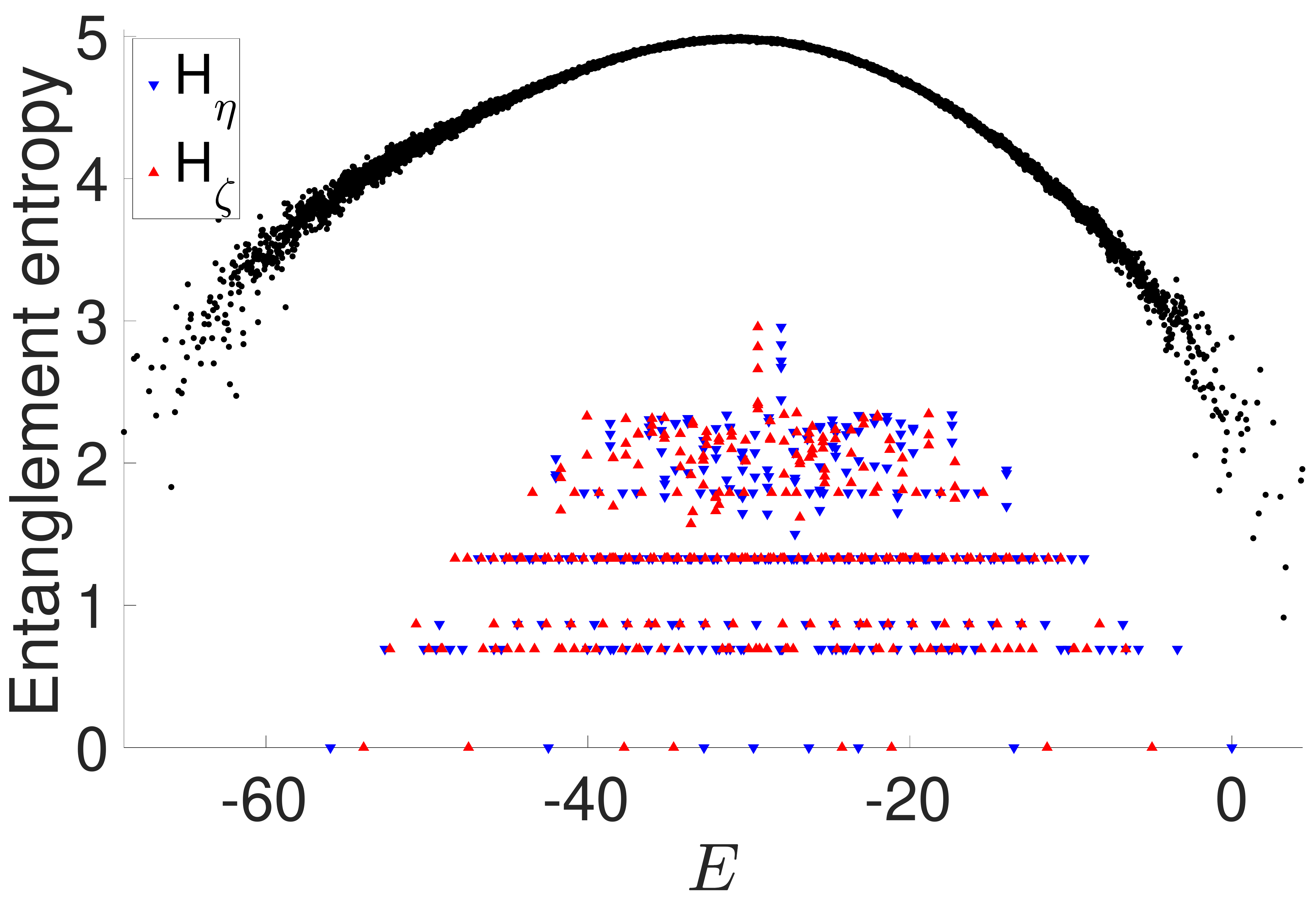}\\
			\includegraphics[width=\columnwidth]{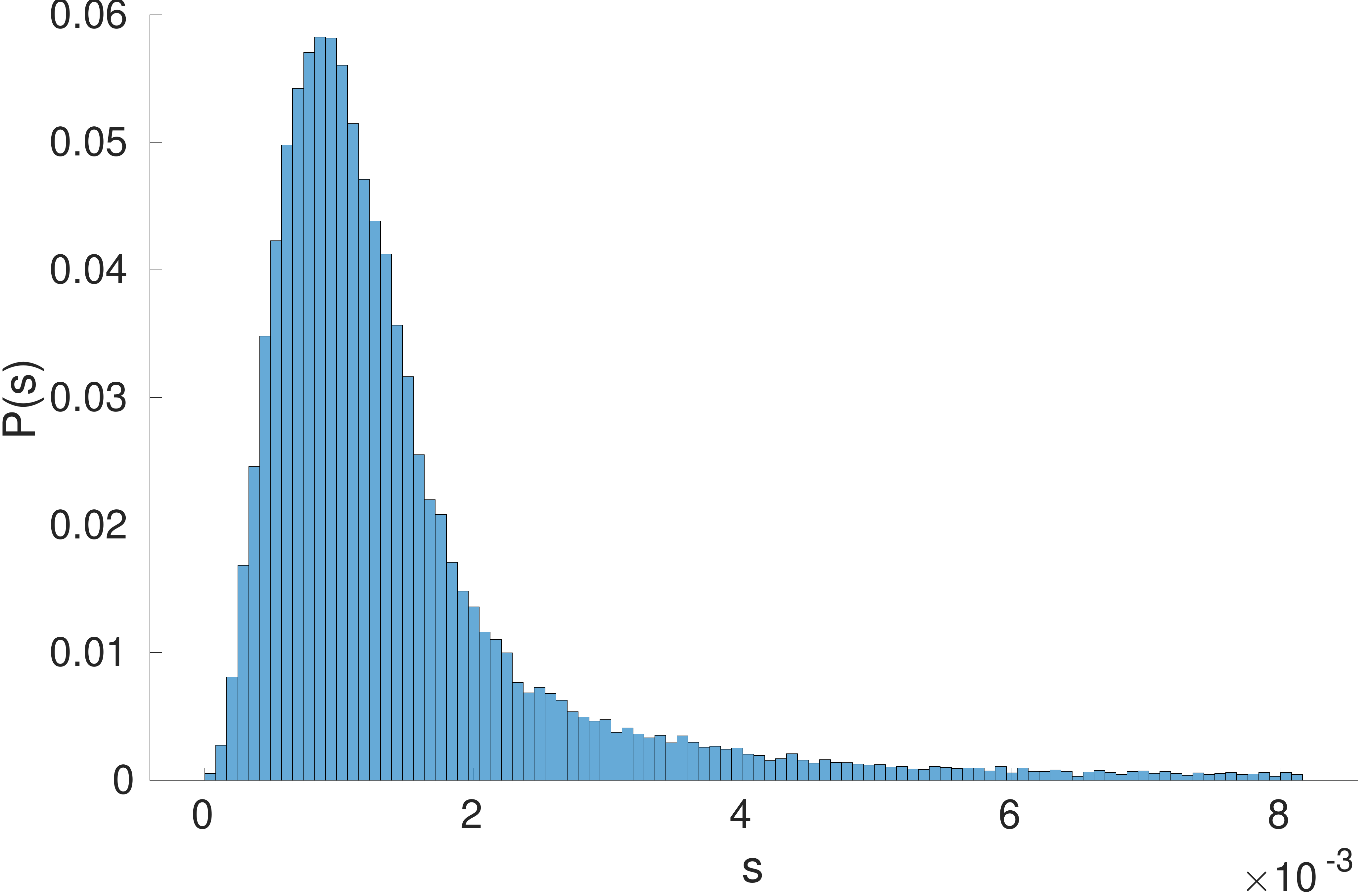}
	\end{center}
\caption{ \label{fig:Mlt4entrAndHist} Numerical results for $N$ = 4, $M$ = 8. Top panel: Entanglement entropy plot for the even sector, with the cut made in the middle of the 1D lattice. There are 16 non-degenerate product scar states with $S=0$ ($8$ in the $\eta$-sector and $8$ in the $\zeta$-sector). Bottom panel: Probability of a level spacing in the even sector of the spectrum. We exclude 3 percent of largest gaps from the plot and from the total norm. }
\end{figure}

The entanglement entropy in every eigenstate in the even fermion parity sector is shown in the top panel of Fig. \ref{fig:Mlt4entrAndHist}. All the 588 $\SO(N)$ singlets 204 of which are (unbreakably) degenerate (64 triple-degenerate, two six-fold degenerate) are significantly less entangled than the generic states at the same temperature.

The average level spacing ratio in the even sector of the fermion number parity is 0.59715 and is approximately the same as the reference value for hermitian random matrices (GUE). Vanishing probability of near-zero gaps that can be seen in the bottom panel of Fig. \ref{fig:Mlt4entrAndHist} indicates level repulsion characteristic of ergodic quantum systems without symmetries.
Combining these observations we can conclude that also in the $M>4$ case the $\SO(N)$ singlets have all the properties of the many-body scars.
We note that with the choice of ``random" $\mu_\alpha$ we made, the system in question has scars (their energy given in eq. \eqref{eq:singletsEnergies}) that are not equally spaced in energy. 

\begin{figure}[htp!]
	\begin{center}
			\includegraphics[width=\columnwidth]{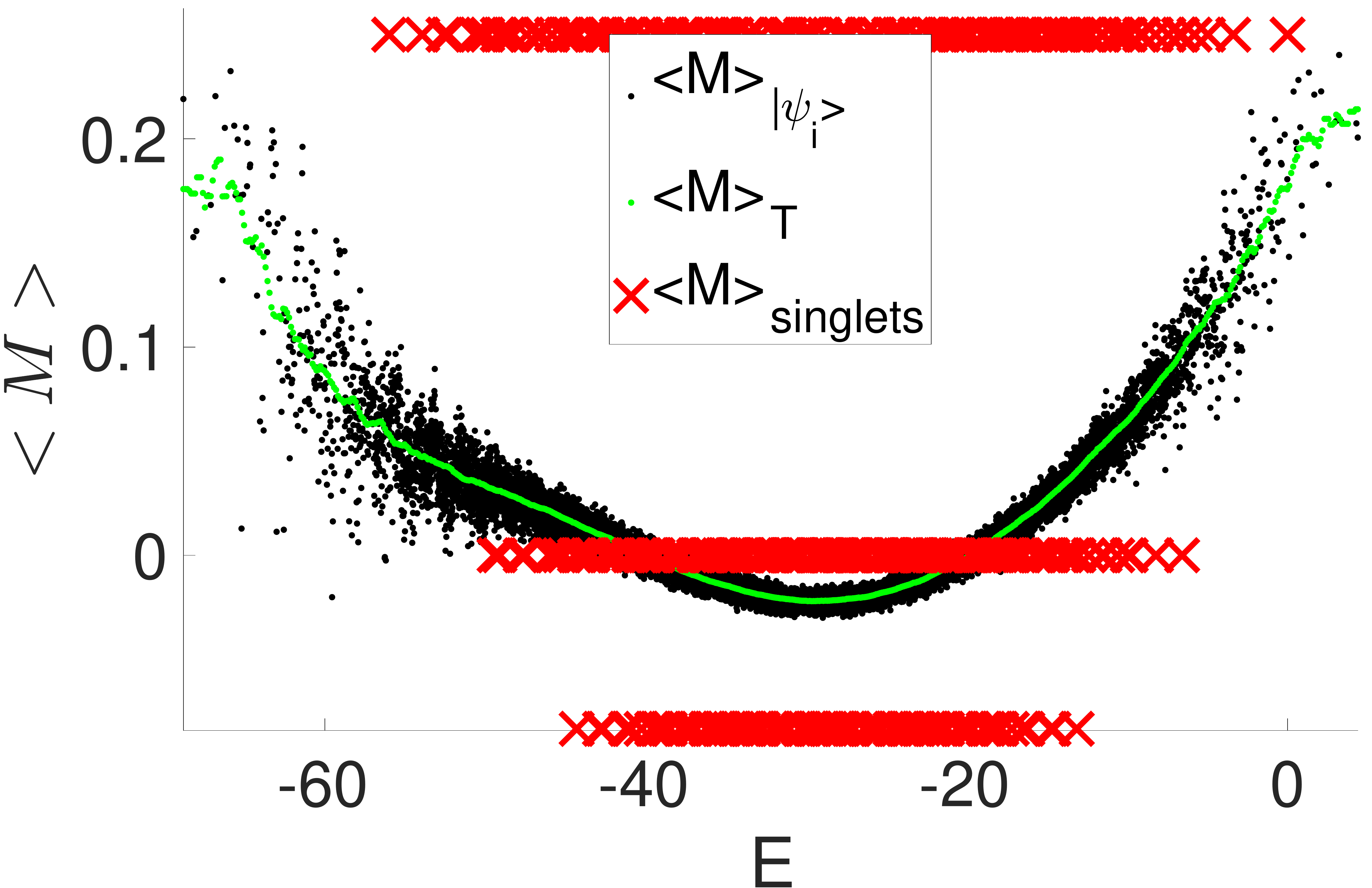}
	\end{center}
\caption{The off-diagonal long-range order \eqref{eq:ODLROdef} with $M=O^{12}_{14}$; measured in every eigenstate (black dots) and in the $\SO(N)$-singlet scar states (red crosses). The green line is the micro-canonical (window) average.
\label{fig:ODLRO} }
\end{figure}

The off-diagonal long-range order (eq. \eqref{eq:ODLROdef}) measured between the most distant sites ($i=1$, $j=4$) for $A=1$ and $B=2$ is shown in Fig. \ref{fig:ODLRO}. As predicted by the eq. \eqref{eq:ODLRO2ndSumRule} the sum of $M=O^{AB}_{14}$ measured in all the scar states normalized by the dimension of the scar subspace equals $\frac{1}{28}$ and is independent of the choice of $A$ and $B$. We confirm numerically that the full minimal set of the ODLRO measurements contains $M-1$ operators \eqref{eq:ODLROdef} where $A$ and $B$ are chosen as $(k,k+1)$ for $k$ between 1 and $M-1$. This means that at least one of the $M-1$ correlators is non-zero in every scar state.

Because of their controlled and predictable values in scars the ODLRO measurements can be used in experiment for detecting invariant many-body scars as an alternative to tracking the projection of the wavefunction on the scar subspace.

\section{Ergodicity breaking in non-local models}
\label{sec:novelFeatures}

The many-body scars reported in the literature so far are typically equidistant in energy. 
Such scars form an integrable subspace. We will now consider an $H_0$ that is non-local and non-integrable by adding the term \eqref{eq:JabJcd} that introduces a strong interaction within the scar subspace. We show that, in this situation, the scar singlets become fully chaotic and ergodic while remaining decoupled from the bulk spectrum.

We investigate the system with $N=4$ and $M=8$ and set the parameters identical to those used for obtaining Figs. \ref{fig:Mlt4entrAndHist} and \ref{fig:ODLRO} but add also the term in eq. \eqref{eq:JabJcd} with the coefficient 0.62598 and the random numbers $r_{ABCD}$ drawn uniformly between 0 and 1.
 This additional term is second order in the generators of $\SO(M)$. It respects the $\SO(N)$ symmetry and is, therefore, of the $H_0$ type. It leaves the scar subspace invariant but mixes up the scars ($\SO(N)$ singlets) and is, therefore, able to generate ergodic spectrum within the scar subspace. Let us also recall that \eqref{eq:JabJcd} breaks the "unbreakable degeneracies" of the singlet states as discussed in Sec. \ref{sec:enSpectrumAndDeg}.

\begin{figure}[htp!]
    \begin{center}
            \includegraphics[width=\columnwidth]{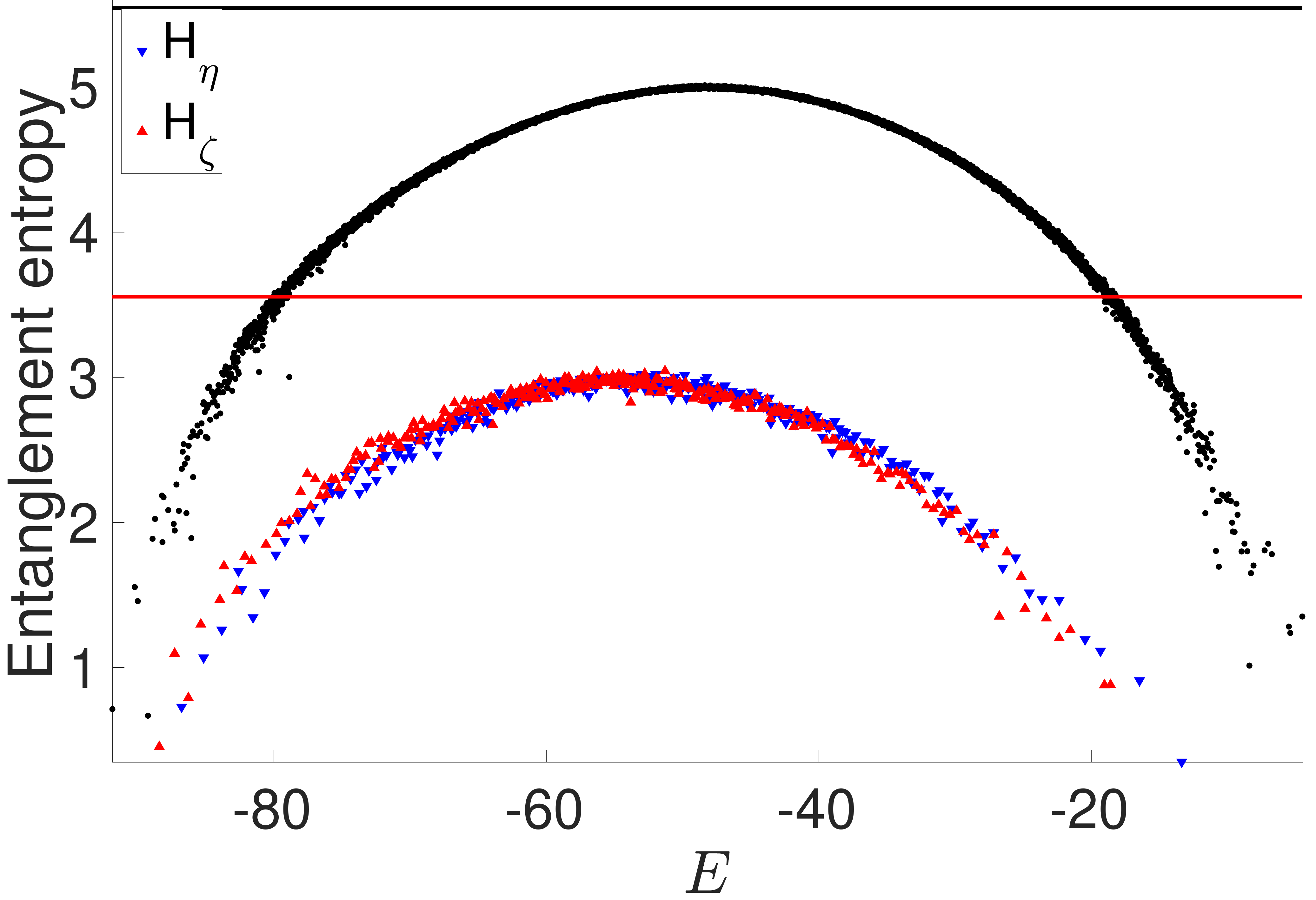}
            \includegraphics[width=\columnwidth]{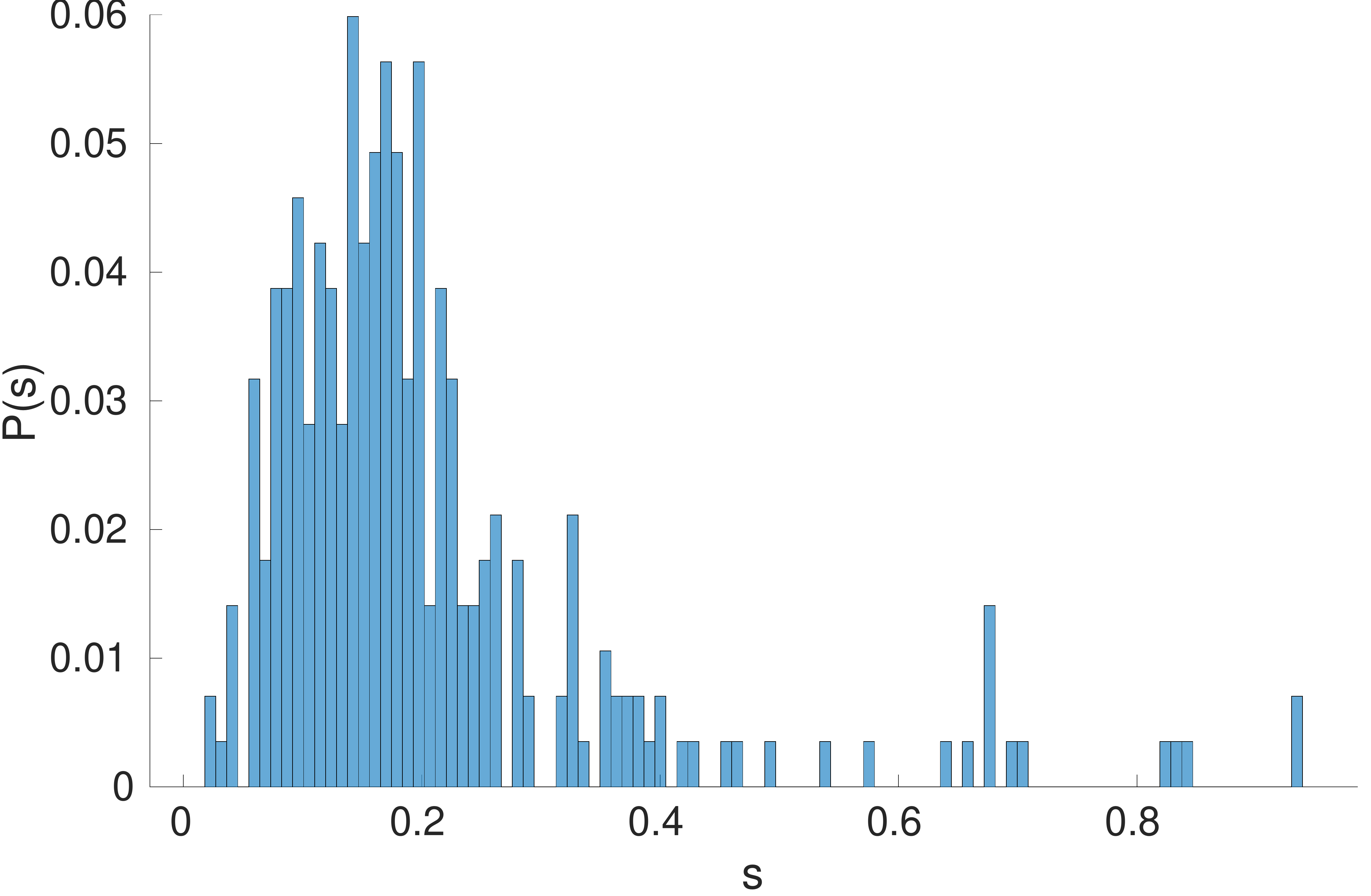}\\       
    \end{center}
\caption{ \label{fig:ergodicScars}
Top panel: 
Entanglement entropy in the even fermionic parity sector for $N=4$, $M=8$ for the Hamiltonian including the scar-mixing term \eqref{eq:JabJcd}. The horizontal lines indicate the bounds on entanglement for a generic and a scar state according to eq. \eqref{entropybound}. 
Bottom panel: 
The level spacing histogram for the 294 states in the $\CH_\eta$ part of the scar subspace. 3 percent of the largest gaps are excluded.
 }
\end{figure}

The resulting entanglement entropies are shown in the top panel of Fig. \ref{fig:ergodicScars}.  The entropies in the scar subspace form a nice thermal arc similarly to the generic states, but separate from the original thermal arc. Note that the entanglement entropy calculation \eqref{Sscale} doesn't apply here because it was performed for the scar wavefunctions that are eigenstates of the simple integrable $H_0$ included in eq. \eqref{eq:fullHNumerics}. Nonetheless, the entanglement entropy should and does satisfy the general bound \eqref{entropybound} shown as the red horizontal line.

In the bottom panel of Fig. \ref{fig:ergodicScars} we show the histogram of the level spacings within the $\CH_\eta$ half of the scar subspace. In spite of the relatively small size of the subspace (294 states), we can see the emergence of the GUE profile (see also the data for a larger system in Fig. \ref{fig:3x6wJab}) with the near-zero gaps being almost absent - an expected signature consequence of level repulsion in an ergodic system without remaining symmetries. The average level statistics ratio $\braket{r}_\eta=0.61801$ (see eq. \eqref{eq:siAndRi}) confirms that the level spacing belongs to the GUE. The level spacing behavior for $\CH_\zeta$ half of the scar subspace (not shown) that is related to $\CH_\eta$ by a symmetry transformation also appears ergodic. The average level spacing there is $\braket{r}_\zeta= 0.60764$. 
 
\section{Enhancing the scar contribution to the density of states \label{sec:dos}}

We now turn to the interesting effects that the scar states in multi-band systems can have on the density of states (DOS). Usually, in the limit of large system size, the scar states form a subspace of measure zero. Therefore, they would not be noticeable in the DOS. This is different in small multi-band systems where a scar subspace can occupy a sizeable fraction of the full Hilbert space. In the two cases $N=2$, $M=16$ and $N=3$, $M=12$ that we investigate numerically, the portion of scars is 
39$\%$ and  
3.2$\%$. This means that, depending on the position of the scar states in energy (that can in principle be controlled, see Sec. \ref{sec:controlScarPos}) the scar states will strongly affect the DOS and could therefore be experimentally detectable. 

\begin{figure}[htp!]
	\begin{center}
			\includegraphics[width=0.49\columnwidth]{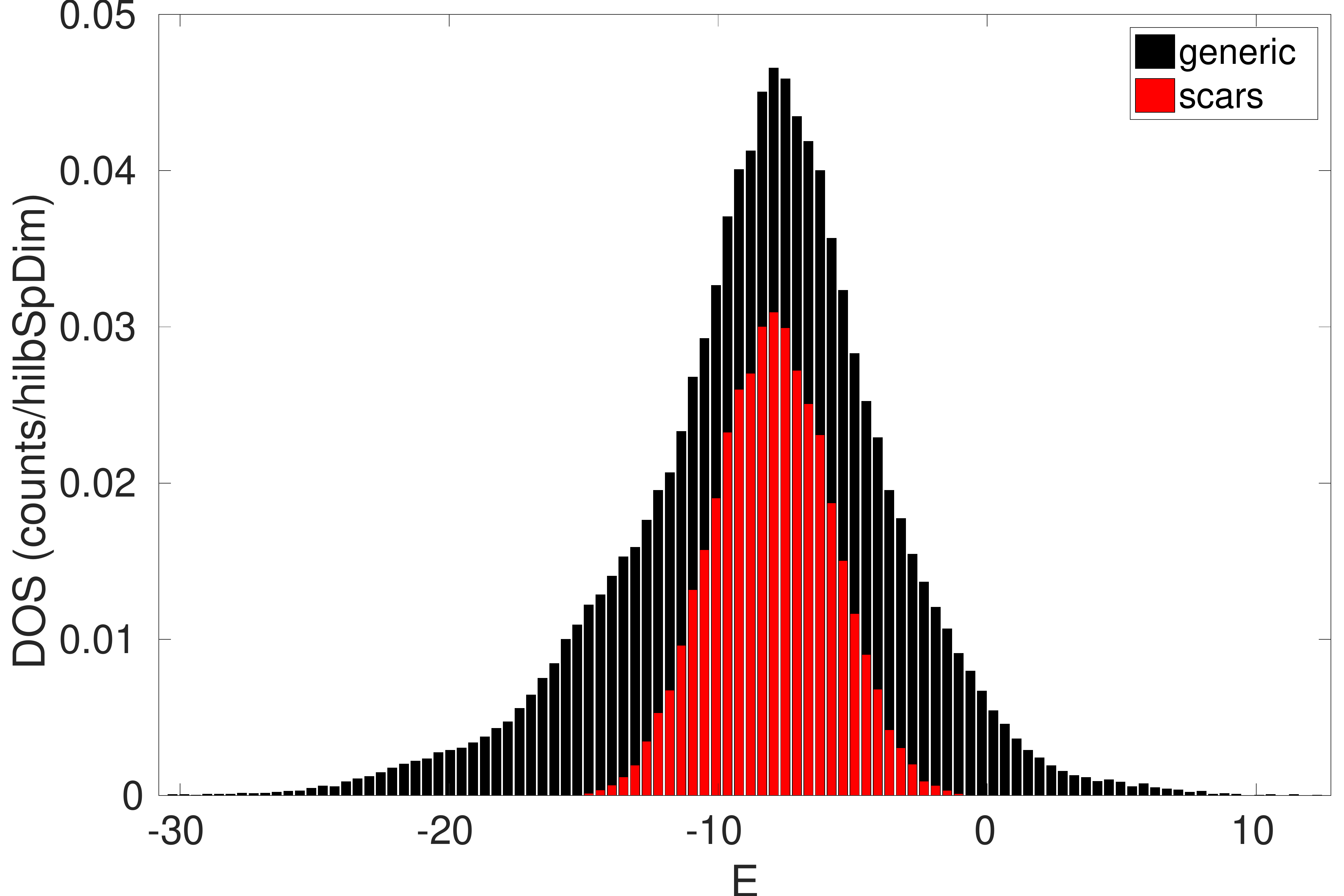}
			\includegraphics[width=0.49\columnwidth]{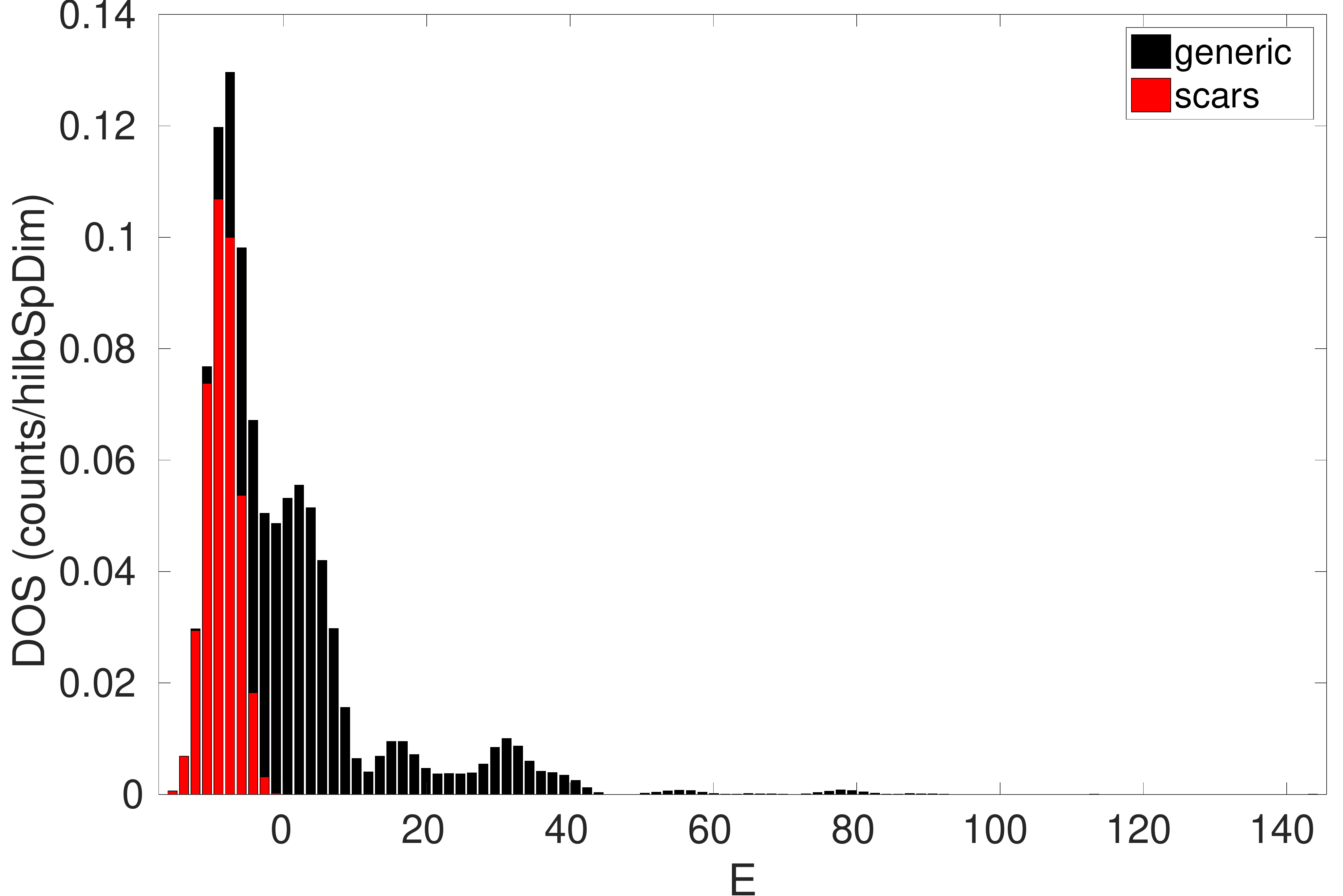}\\
			\includegraphics[width=0.49\columnwidth]{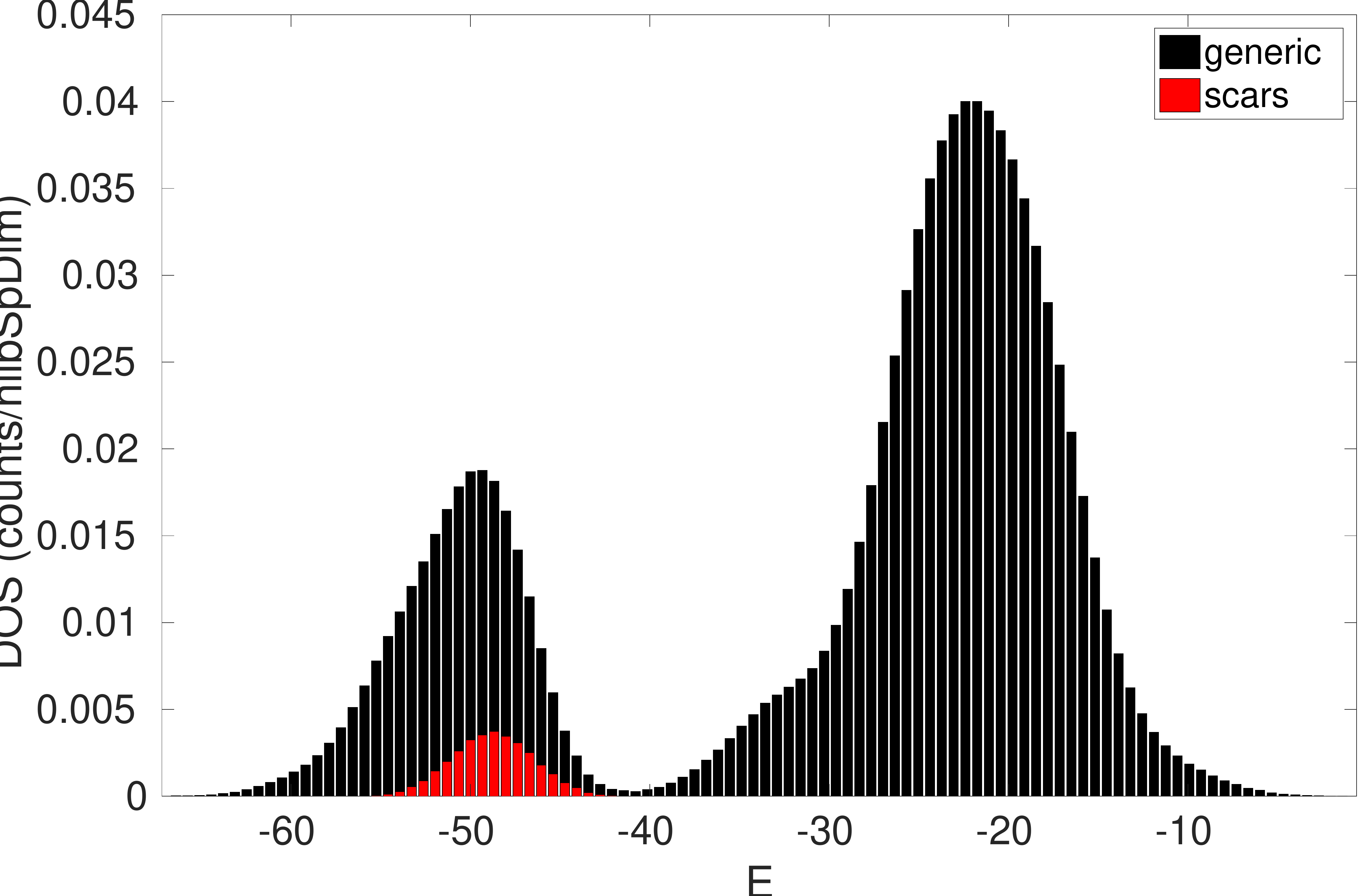}
			\includegraphics[width=0.49\columnwidth]{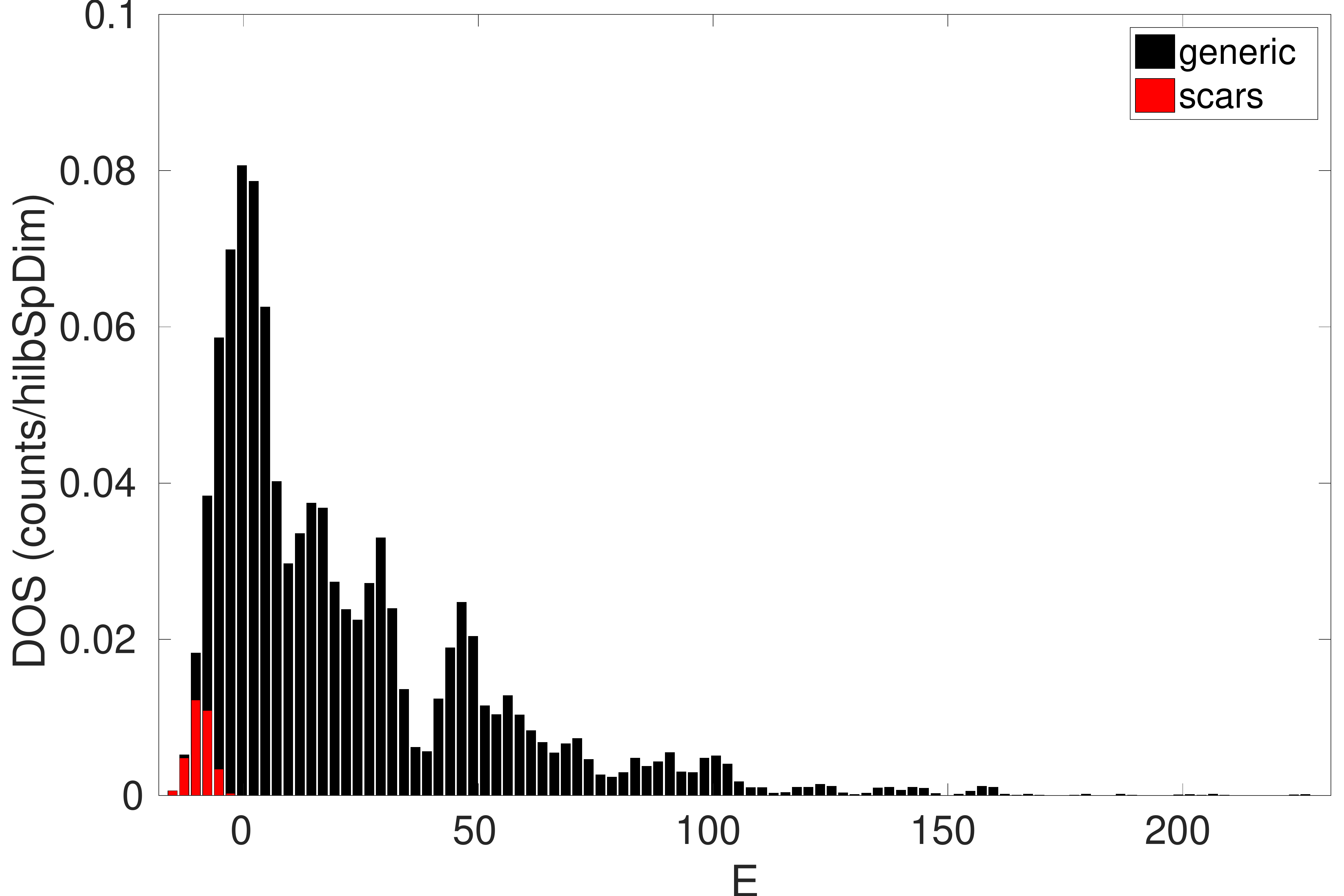}
	\end{center}
\caption{ \label{fig:scarsDOS}
The density of states where the contribution of scars in every energy window is shown in red. The Hamiltonian is \eqref{eq:fullHNumerics} and the parameters are the same as in Fig. \ref{fig:Mlt4entrAndHist} except that all the chemical potentials are scaled down by a factor of 2/9. Further modifications are indicated below. Top: $N=2$, $M=16$, even fermionic parity sector (contains both $\eta$ and $\zeta$ scars because $N$ is even) Top Left: no further modifications to the Hamiltonian. Top Right: the positive-definite local term \eqref{eq:HUTsq} with the coefficient $4.52691$
 is added. Bottom: $N=3$, $M=12$. The DOS is shown in the odd fermionic parity sector (contains only the $\zeta$ scars because $N$ is odd). Bottom Left: the Hamiltonian is modified by scaling up the $U$ interaction term by a factor of 150. Bottom Right: the term \eqref{eq:HUTsq}  is added with the strength of $9.05382$.
}
\end{figure}

Fig. \ref{fig:scarsDOS} shows the density of states with the scars contribution indicated separately for the systems with $N=2$, $M=16$ and $N=3$, $M=12$. Scars make a significant effect on the shape of the DOS in certain energy ranges (we do, however, expect this effect to weaken if we increase $N$ while keeping $M$ fixed). Using the term \eqref{eq:HUTsq} the scars can also be localized near the low-energy part of the spectrum. The prevalence of scars in DOS could be seen in any measurement made at the corresponding temperature. This greatly simplifies the experimental studies of scars eliminating the need to prepare a specific initial state of the system. We emphasize that in all the cases presented in Fig. \ref{fig:scarsDOS} the full Hamiltonian is local. 
Further possibilities of engineering the shape of the DOS  are illustrated in the Fig. \ref{fig:scarsDOSapp} in the appendix.

The dimension of the scar subspace \eqref{eq:gendimfor} quickly grows with $N$ and $M$. The full Hilbert space dimension however grows even faster. Therefore the exotic signatures of scars in the DOS can only be observed in small systems. If we qualitatively set a $0.1\%$ threshold on the fraction of the scars in the Hilbert space then the largest suitable systems are $(N,M):(6,6),(4,24),(3,40)$. While small from the real material perspective they may be well suited for the existing quantum simulators based on cold atoms or quantum computing devices.

\section{Discussion}

We have presented the structure of the quantum many-body scars in lattice systems of $N$ sites with $M$ Majorana fermions per site. Following the idea of group-invariant scars \cite{pakrouski2020GroupInvariantScars,Pakrouski:2021jon} we identified the classes of ergodic Hamiltonians where the $\SO(N)$ invariant states are exact eigenstates and many-body scars. 
Analytical expressions for the energies of scars and their signature off-diagonal long-range order correlators are provided. We specify how the scar wavefunctions can be built for arbitrary $M$ and provide their explicit wave functions for $M=6$.
Some of these generalized $\eta$-pairing states were found also in \cite{Nakagawa:2022jsg}.

The upper bound on the entanglement entropy of scars is derived for arbitrary $M$. It grows logarithmically with the sub-system size, generalizing the earlier results for $M=4$ \cite{Vafek_2017} and $M=6$ \cite{Nakagawa:2022jsg}. Furthermore, we obtain an upper bound on the entanglement entropy for any group-invariant states, not limited to the multi-flavor Majorana Hilbert spaces considered in this work. This leads us to a general conclusion that any group-invariant many-body scars must always have the entanglement entropy that is parametrically lower than that of generic states. This is in agreement with the calculations in almost all models with many-body scars, where the general argument for the reduced entropy has not been provided yet.

The Hilbert space of multi-flavor Majorana fermions used in this work allowed us to uncover several possibilities for the behavior of scars that were not discussed in the earlier literature.
The number of states breaking ergodicity in a system with $N$ sites and $M$ flavors grows as $N^{M(M-2)/8}$.  
This means that for some system sizes the scars can occupy a sizeable fraction of the Hilbert space and can be clearly seen in the density of states.
For $M>6$ we find degeneracies in the scar subspace that cannot be lifted by the local interactions which preserve the decoupling of scars. These degeneracies present a new promising resource that could potentially be used for robust quantum computing, similarly to how the topological degeneracies are used in topological quantum computing schemes \cite{freedman2003topological}.

Although the many-body scars we study can exhibit revivals under some conditions, they are not in general equally-spaced in energy. 
Further, we demonstrated that by considering a non-local interaction the spectrum of scars can be made ergodic.
It is an interesting question for the future studies if this requirement applies to all many-body scars in general. Should systems with ergodic scars be implemented in local systems, the experimental consequences of the simultaneous presence of two distinct thermal averages (as in Fig. \ref{fig:ergodicScars}) is another intriguing direction.

Finally, the models with multiple Majorana flavors possess a rich variety of the off-diagonal long-range order two-point correlators. The result of measuring such correlators in a scar state does not depend on the distance between the points. Similarly to the superconducting correlations found in the $\eta$ states for $M=4$, the full set of the corresponding ODLRO operators we identified for $M>4$, may be used for detecting many-body scars experimentally. 

\section*{Acknowledgements}

We thank T. Iadecola, O. Motrunich, S. Moudgalya, Z. Papi\' c and G. Tarnopolsky for useful discussions.
The simulations presented in this work were performed on computational resources managed and supported by Princeton's Institute for Computational Science $\&$ Engineering and OIT Research Computing.
This research was supported in part by the US NSF under Grant No.~PHY-2209997 and by the Princeton Gravity Initiative. 
F.K.P. is currently a Simons Junior Fellow at NYU and supported by a grant 855325FP from the Simons Foundation. Some of this research was carried out while I.R.K. was on a sabbatical leave, and he is grateful to the Institute for Advanced Study for hosting him during the 2021-22 academic year.

\newpage

\appendix 

\section{A brief review of $\so(2n,\mathbb C)$ and its representations}\label{SOreview}
The complexified Lie algebra $\so(2n,\mathbb C)$ are spanned by all antisymmetric matrices over $\mathbb C$. A convenient basis of $\so(2n,\mathbb C)$ is given by
\begin{align}
J^{AB}=E^{AB}-E^{BA},\quad (E^{AB})_{CD}=\delta_{C}^A\delta_{D}^B
\end{align}
where $E^{AB}$ is an $2n\times 2n$ matrix with only one non-zero entry. $J^{AB}$ constructed in this way satisfy the commutation relation (\ref{JJ}), and they are realized as anti-hermitian operators acting on certain Hilbert space in a unitary representation. The standard definition of quadratic Casimir is 
\begin{align}\label{Casdefo}
\CC_2^{\SO(2n)}=\frac{1}{2}\sum_{A,B=1}^{2n}\,J^{AB}J^{BA}
\end{align}
Choose Cartan generators to be
\begin{align}
h_\alpha=-i J^{2\alpha-1, 2\alpha}, \,\,\,\,\, \alpha=1,2,\cdots, n
\end{align}
and they span a Cartan subalgebra $\mathfrak{h}$. 
Given this choice of Cartan subalgebra, positive roots are $e_\alpha\pm e_\beta, \,\,\,\, 1\le \alpha<\beta\le n$, with $\{e_\alpha\}$ being the standard basis on $\mathbb R^n$. The corresponding $\SO(2n)$ generators  are 
\small
\begin{align}
&e_\alpha\!-\!e_\beta: \zeta^\dagger_{\alpha\beta}=\frac{J^{2\alpha\!-\!1, 2\beta\!-\!1}\!-\!iJ^{2\alpha\!-\!1, 2\beta}\!+\!i \left(J^{2\alpha, 2\beta\!-\!1}\!-\!i J^{2\alpha, 2\beta}\right)}{2}\nonumber\\
&e_\alpha\!+\!e_\beta: \eta^\dagger_{\alpha\beta}=\frac{J^{2\alpha\!-\!1, 2\beta\!-\!1}\!+\!i J^{2\alpha\!-\!1, 2\beta}\!+\!i \left(J^{2\alpha, 2\beta\!-\!1}\!+\!i J^{2\alpha, 2\beta}\right)}{2}
\end{align}
\normalsize
where the overall normalization factor $\frac{1}{2}$ is inserted such that $[\zeta^\dagger_{\alpha\beta},\eta_{\beta\gamma}^\dagger]=\eta_{\alpha\gamma}^\dagger$.
The generators corresponding to negative roots $-(e_\alpha\pm e_\beta)$ are hermitian conjugate of $\zeta^\dagger_{\alpha\beta}$ and $\eta^\dagger_{\alpha\beta}$. Altogether, the root decomposition of $\so(2n,\mathbb C)$ is 
\small
\begin{align}
\so(2n,\mathbb C)=\mathfrak{h}\bigoplus\bigoplus_{1\le \alpha<\beta\le n} \left(\mathbb{C}\zeta^\dagger_{\alpha\beta}\oplus \mathbb{C}\eta^\dagger_{\alpha\beta}\oplus \mathbb{C}\zeta_{\alpha\beta}\oplus \mathbb{C}\eta_{\alpha\beta}\right)
\end{align}
\normalsize

An integral highest-weight vector $\lambda$ can be parameterized by $\lambda=\sum_{\alpha=1}^n \lambda_\alpha e_\alpha$, where $\lambda_\alpha$ are either integers or half-integers, satisfying $\lambda_1\ge\lambda_2\ge\cdots\ge \lambda_{n-1}\ge |\lambda_n|$. In terms of Young diagram, $\lambda_\alpha$ is the number of boxes in the $\alpha$-th row.  Given a highest-weight vector $\lambda=(\lambda_1,\cdots,\lambda_n)$, the Casimir $\CC_2^{\SO(2n)}$ defined by eq. (\ref{Casdefo}) takes the value 
\begin{align}\label{SOcas}
\CC_2^{\SO(2n)}(\lambda)=\sum_{\alpha=1}^n\lambda_\alpha(\lambda_\alpha+2n-2\alpha)
\end{align}
which cannot be used to distinguish the two highest-weight representations $(\lambda_1,\cdots,\pm\lambda_n)$. For a rectangular Young diagram $\lambda=(k^n)$ ($k$ is an arbitrary nonnegative integer), eq.(\ref{SOcas}) yields 
\begin{align}\label{specas}
\CC_2^{\SO(2n)}(k^n)=\CC_2^{\SO(2k)}(n^k)=k n(n+k-1)
\end{align}
The Weyl character of the $\lambda$-representation, defined as the trace of $x_1^{h_1}\cdots x_n^{h_n}$ over the Hilbert space, is given by 
\begin{align}\label{Weylchar}
 \chi^{\SO(2n)}_{\lambda} (\bm x)
 &=\frac{\det\left(x_\alpha^{\ell_\beta}+x_\alpha^{-\ell_\beta}\right)+\det\left(x_\alpha^{\ell_\beta}-x_\alpha^{-\ell_\beta}\right)}{\det\left(x_\alpha^{n-\beta}+x_\alpha^{-(n-\beta)}\right)}
\end{align}
where $\ell_\alpha=n+\lambda_\alpha-\alpha$.
Taking the $x_\alpha\to 1$ limit in $ \chi^{\SO(2n)}_{\lambda} (\bm x)$ yields the dimension of this representation
\begin{align}\label{SOdim}
\dim^{\SO(2n)}_\lambda=\prod_{\alpha<\beta}\frac{(\ell_\alpha-\ell_\beta)(\ell_\alpha+\ell_\beta)}{(\beta-\alpha)(2n-\alpha-\beta)}
\end{align}

\section{Additional numerical results }\label{sec:apNum}

\begin{figure}[htp!]
	\begin{center}
			\includegraphics[width=\columnwidth]{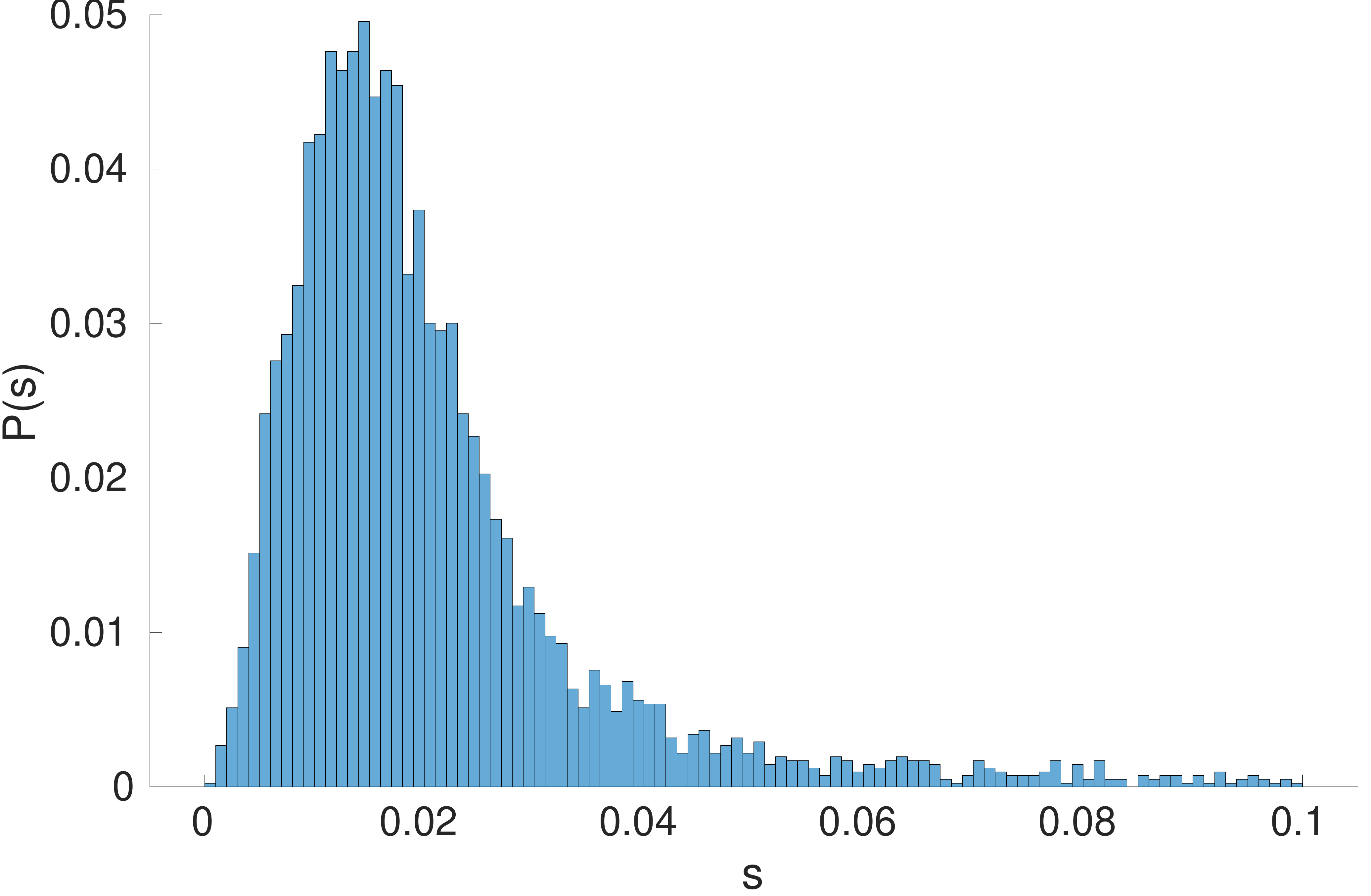}
	\end{center}
\caption{ \label{fig:3x6wJab} $N=3$, $M=12$. Histogram of the energy gaps within the $\CH_\eta$ subspace in even fermion parity sector. 3 percent of largest gaps excluded.}
\end{figure}

Fig. \ref{fig:3x6wJab} shows the distribution of the energy gaps in the scar subspace for the model parameters that are identical to those used for Fig. \ref{fig:ergodicScars} but in a larger system with 4224 $\eta$ scars.
The average level statistics ratio here is $\braket{r}_\eta=0.59335$ and $\braket{r}_\zeta=0.60095$ for the $\zeta$ states (histogram is qualitatively the same but not shown). As in the main text the data is fully consistent with an ergodic spectrum without remaining symmetries also within the scar subspace.  

\begin{figure}[htp!]
	\begin{center}
			\includegraphics[width=\columnwidth]{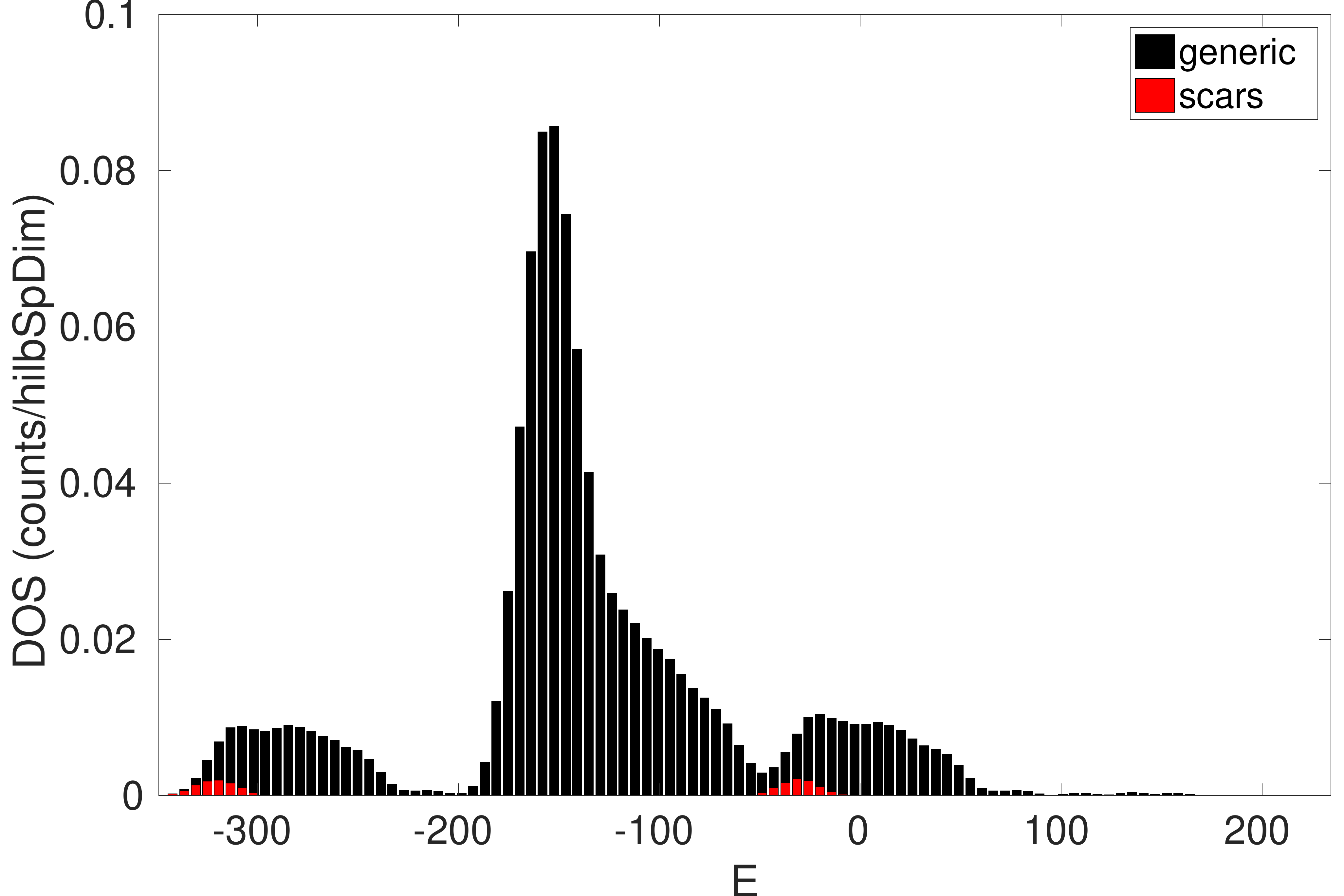}
	\end{center}
\caption{ \label{fig:scarsDOSapp} $N=4$, $M=8$. Even fermionic parity sector. Hamiltonian is the same as in Fig. \ref{fig:Mlt4entrAndHist} but the $U$ interaction term is scaled up by a factor of 200 and the positive-definite term \eqref{eq:HUTsq} is added with strength 11.31727.}
\end{figure}

Fig. \ref{fig:scarsDOSapp} illustrates that by combining the stronger U interaction and the positive-definite term \eqref{eq:HUTsq} the scars can be exposed even in the DOS of the systems where they only occupy a vanishingly small part of the Hilbert space.

\section{Entanglement entropy for $M=6$}
\label{sec:m6entropy}

Using the explicit expression (\ref{normalscar}) for all the $\eta$-states in the $M=6$ case, we can calculate their entanglement entropy analytically.
We divide the underlying lattice into two disjoint subsets $\Sigma_1$ and $\Sigma_2$. For example, $\Sigma_1$ consists of the first $N_1$ sites, i.e. $i=1,2,\cdots, N_1$, and $\Sigma_2$ consists of the rest $N_2=N-N_1$ sites. In each sublattice $\Sigma_a$, there is an empty vacuum $|0\rangle_a$ satisfying $|0\rangle = |0\rangle_1\otimes|0\rangle_2$.
Then we are allowed  to split the $\eta$-operators in the following way
\begin{gather}
\eta^{1\dagger}_{\alpha\beta} = \sum^{N_1}_{i=1}c^\dagger_{i\alpha} c^\dagger_{i\beta}, \quad \eta^{2\dagger}_{\alpha\beta} = \sum^{N}_{i=N_1+1}c^\dagger_{i\alpha} c^\dagger_{i\beta}, \notag
\end{gather} 
such that $\eta^{a\dagger}_{\alpha\beta}$ can excite $\eta$ states of the subsystem $\Sigma_a$ on $|0\rangle_a$
\begin{align}
|m_{12}, m_{13},m_{23}\rangle_a=C_{\bm m}(N_a)\prod_{\alpha<\beta}(\eta^{a\dagger}_{\alpha\beta})^{m_{\alpha\beta}}|0\rangle_a .
\end{align}
Because of $\eta^\dagger_{\alpha\beta} = \eta^{1\dagger}_{\alpha\beta} + \eta^{2\dagger}_{\alpha\beta}$, we have a tensor product decomposition of any $\eta$ state $|\bm k\rangle\equiv |k_{12}, k_{13}, k_{23}\rangle$ defined in eq. (\ref{normalscar})
\small
\begin{gather}
\ket{\bm k } = C_{\bm k}(N)\sum_{m_{\alpha\beta}=0}^{k_{\alpha\beta}} \prod_{\alpha<\beta} \binom{k_{\alpha\beta}}{m_{\alpha\beta}} \left(\eta^{1\dagger}_{\alpha\beta}\right)^{m_{\alpha\beta}} \left(\eta^{2\dagger}_{\alpha\beta}\right)^{k_{\alpha\beta}-m_{\alpha\beta}} \ket{0}  \notag\\
= \sum^{k_{\alpha\beta}}_{m_{\alpha\beta}=0}\frac{C_{\bm k}(N)}{C_{\bm m}(N_1) C_{\bm k-\bm m}(N_2)}\prod_{\alpha<\beta}\binom{k_{\alpha\beta}}{m_{\alpha\beta}} \ket{\bm m}_1 \otimes \ket{\bm k-\bm m}_2\ . \notag \label{eq:decom}
\end{gather}
\normalsize
Let us note that $\ket{\bm m}_1$ and $\ket{\bm k - \bm m}_2$ are also singlet states in each subsector.
Taking the partial trace over the Hilbert space of $\Sigma_2$ yields the reduced density matrix $\rho_{\Sigma_1}(\bm k)$ of $|\bm k \rangle$
\begin{gather}\label{jll}
\rho_{\Sigma_1}(\bm k)=\sum^{k_{\alpha\beta}}_{m_{\alpha\beta}=0} \lambda_{\bm k}( \bm m)\ket{\bm m}_1 \bra{\bm m}_1,
\end{gather}
where
\small
\begin{align}
\label{eq:lambdaK}
&\lambda_{\bm k}(\bm m)=\frac{C^{N_1}_{m_{12}, m_{13}, m_{23}}C^{N_2}_{k_{12}-m_{12}, k_{13}-m_{13}, k_{23}-m_{23}}}{C^N_{k_{12}, k_{13}, k_{23}}}\nonumber\\
& C^{N}_{a,b,c}\equiv \frac{N!}{a!\,b!\,c!\,(N-a-b-c)!}
\end{align}
\normalsize
and $\lambda_{\bm k}(\bm m)$ vanishes when $m_T> N_1$ or $k_T-m_T> N_2$.  The density matrix $\rho_{\Sigma_1}$ corresponds to a pure state if (i) all $k_{\alpha\beta}$ are vanishing, or (ii) one $k_{\alpha\beta}$ is equal to $N$ and the rest are vanishing. The former is trivial since it implies $m_{12}=m_{13}=m_{23}=0$. For the latter, say $k_{12}=N$, we have first $m_{13}=m_{23}=0$. Then nonvanishing  $\lambda_{\bm k}(\bm m)$ requires $m_T\ge k_T-N_2=N_1$ and $m_T\le N_1$, which completely fix $m_{12}=N_1$. Indeed, (i) corresponds to $|0\rangle$, and (ii) corresponds to $\CA^\dagger_\alpha \CA^\dagger_\beta|0\rangle, 1\le \alpha<\beta\le 3$, which are the only product states in $\CH_\eta$.

Next we proceed to compute the entanglement entropy of $\rho_{\Sigma_1}(\bm k)$ in the thermodynamic limit, defined as the limit of  $N\to\infty, k_{\alpha\beta}\to \infty$ such that  $\nu_{\alpha\beta}\equiv \frac{k_{\alpha\beta}}{N}$ are finite.  
We further choose $N_1\ll N$ so that $\Sigma_2$ can be treated as a heat bath and meanwhile keep $N_1\gg 1$ to allow scaling of entanglement entropy $\rho_{\Sigma_1}$.  
In this limit, $\lambda_{\bm k}(\bm m)$ is sharply peaked around $m_{\alpha\beta}=m_{\alpha\beta}^*\equiv \nu_{\alpha\beta} N_1$.
At this extremal point, $\lambda(\bm m^*)\approx \frac{1}{\sqrt{(2\pi N_1)^3\nu_{12}\nu_{13}\nu_{23}(1-\nu_T)}}$ with $\nu_T=\nu_{12}+\nu_{13}+\nu_{23}$.  For general $\bm m$, the matrix element $\lambda(\bm m)$ is approximated by a 3D Gaussian centered at $\bm m^*$
\begin{align}
\lambda(\bm m)\approx  \frac{e^{-\frac{1}{2 N_1} (\bm m-\bm m^*)^T \mathcal {M} (\bm m-\bm m^*)}}{\sqrt{(2\pi N_1)^3\nu_{12}\nu_{13}\nu_{23}(1-\nu_T)}}
\end{align}
where $\mathcal M$ is a $3\times 3$ symmetric matrix, given by 
\begin{align}
\mathcal M=\begin{pmatrix}\frac{1}{\nu_{12}}+\frac{1}{1-\nu_T} & \frac{1}{1-\nu_T} & \frac{1}{1-\nu_T}\\
\frac{1}{1-\nu_T} &\frac{1}{\nu_{13}}+ \frac{1}{1-\nu_T} & \frac{1}{1-\nu_T}\\
\frac{1}{1-\nu_T} & \frac{1}{1-\nu_T} &\frac{1}{\nu_{23}}+ \frac{1}{1-\nu_T}\\
\end{pmatrix}
\end{align}
with determinant equal to $\nu_{12}\nu_{13}\nu_{23}(1-\nu_T)$. The  scaling property of the entanglement entropy of $\rho_{\Sigma_1}(\bm k)$ in thermodynamic limit can then be computed by replacing the sum over $\bm m$ with a triple integral $\int d^3\bm m$: 
\begin{align}
&S_{\Sigma_1}(\bm k)\approx-\int d^3\bm m \,\lambda_{\bm k}(\bm m)\, \log \lambda_{\bm k} (\bm m)\nonumber\\
&\approx \frac{1}{2}\log\left((2\pi N_1)^3\nu_{12}\nu_{13}\nu_{23}(1-\nu_T)\right)\sim\frac{3}{2}\log (N_1)\label{Sscale}
\end{align}

The calculation of entanglement entropy works analogously for the other $\SO(N)$ invariant subspace $\CH_\zeta$.  In particular, starting with a $\zeta$ state $|\bm k\rangle^\zeta$, we end up with same density matrix (\ref{jll}), with $|\bm m\rangle_1$ being replaced by the corresponding $\zeta$ states on $\Sigma_1$.

\bibliography{scar}

\end{document}